\documentclass[twocolumn,aps,prb,10pt,superscriptaddress,longbibliography]{revtex4-2}
\usepackage{amsmath}
\usepackage{bm}
\usepackage{glossaries}
\usepackage{graphicx}
\usepackage{multirow}
\usepackage{threeparttable}
\usepackage[usenames,dvipsnames,svgnames]{xcolor}
\usepackage{hyperref}
\hypersetup{
pdfnewwindow=true,      
colorlinks=true,        
linkcolor=Blue,         
citecolor=Blue,         
filecolor=Blue,         
urlcolor=Blue           
}
\usepackage{siunitx}
\newacronym{mof}{MOF}{metal-organic framework}
\newacronym{zif}{ZIF}{zeolitic imidazolate framework}
\newacronym{dft}{DFT}{density functional theory}
\newacronym{md}{MD}{molecular dynamics}
\newacronym{emp}{EMP}{empirical potential}
\newacronym{mlp}{MLP}{machine-learned potential}
\newacronym{gap}{GAP}{Gaussian approximation potential}
\newacronym{nep}{NEP}{neuroevolution potential}
\newacronym{reax}{ReaxFF}{reactive force field}
\newacronym{uff}{UFF}{Universal force field}
\newacronym{mofff}{MOF-FF}{force field for MOFs}
\newacronym{btwff}{BTW-FF}{Bristow-Tiana-Walsh force field}
\newacronym{zifff}{ZIF-FF}{force field for ZIFs}
\newacronym{nn}{NN}{neural network}
\newacronym{snes}{SNES}{separable natural evolution strategy}
\newacronym{mbd}{MBD}{many-body dispersion}
\newacronym{rmse}{RMSE}{root mean square error}
\newacronym{hnemd}{HNEMD}{homogeneous non-equilibrium molecular dynamics}
\newacronym{nemd}{NEMD}{non-equilibrium molecular dynamics}
\newacronym{emd}{EMD}{equilibrium molecular dynamics}
\newacronym{shc}{SHC}{spectral heat current}
\newacronym{mfp}{MFP}{mean free path}
\newacronym{bte}{BTE}{Boltzmann transport equation}
\newacronym{ald}{ALD}{anharmonic lattice dynamics}
\newacronym{aimd}{AIMD}{\emph{ab initio} molecular dynamics}
\newacronym{pc}{PC}{principal component}
\newacronym{2d}{2D}{two-dimensional}
\newacronym{sed}{SED}{spectral energy density}
\newacronym{rdf}{RDF}{radial distribution function}
\newacronym{adf}{ADF}{angular distribution function}
\newacronym{si}{SI}{Supplemental Information}
\newacronym{ltc}{LTC}{lattice thermal conductivity}
\newacronym{hac}{HAC}{heat current autocorrelation}

\DeclareSIUnit\angstrom{\text{Å}}
\DeclareSIUnit{\atom}{atom}
\DeclareSIUnit{\step}{step}
\DeclareSIUnit{\atomstepsecond}{\atom\step\per\second}

\begin{document}

\title{Sub-micrometer phonon mean free paths in metal-organic frameworks revealed by machine-learning molecular dynamics simulations}

\author{Penghua Ying}
\email{hityingph@163.com}
\affiliation{School of Science, Harbin Institute of Technology, Shenzhen, 518055, P. R. China}

\author{Ting Liang}
\affiliation{Department of Electronic Engineering and Materials Science and Technology Research Center, The Chinese University of Hong Kong, Shatin, N.T., Hong Kong SAR, 999077, P. R. China}

\author{Ke Xu}
 \affiliation{Department of Physics, Xiamen University, Xiamen 361005, P. R. China.}

\author{Jin Zhang}
\affiliation{School of Science, Harbin Institute of Technology, Shenzhen, 518055, P. R. China}

\author{Jianbin Xu}
\affiliation{Department of Electronic Engineering and Materials Science and Technology Research Center, The Chinese University of Hong Kong, Shatin, N.T., Hong Kong SAR, 999077, P. R. China}

\author{Zheng Zhong}
\affiliation{School of Science, Harbin Institute of Technology, Shenzhen, 518055, P. R. China}

\author{Zheyong Fan}
\email{brucenju@gmail.com}
\affiliation{College of Physical Science and Technology, Bohai University, Jinzhou 121013, P. R. China}

\date{\today}

\begin{abstract} 
Metal-organic frameworks (MOFs) are a family of materials that have high porosity and structural tunability and hold great potential in various applications, many of which requiring a proper understanding of the thermal transport properties. Molecular dynamics (MD) simulations play an important role in characterizing the thermal transport properties of various materials. However, due to the complexity of the structures, it is difficult to construct accurate empirical interatomic potentials for reliable MD simulations of MOFs. To this end, we develop a set of accurate yet highly efficient machine-learned potentials for three typical MOFs, including MOF-5, HKUST-1, and ZIF-8, using the neuroevolution potential approach as implemented in the GPUMD package, and perform extensive MD simulations to study thermal transport in the three MOFs. Although the lattice thermal conductivity (LTC) values of the three MOFs are all predicted to be smaller than 1 $\rm{W/(m\ K)}$ at room temperature, the phonon mean free paths (MFPs) are found to reach the sub-micrometer scale in the low-frequency region. As a consequence, the apparent LTC only converges to the diffusive limit for micrometer single crystals, which means that the LTC is heavily reduced in nanocrystalline MOFs. The sub-micrometer phonon MFPs are also found to be correlated with a moderate temperature dependence of LTC between those in typical crystalline and amorphous materials. Both the large phonon MFPs and the moderate temperature dependence of LTC fundamentally change our understanding of thermal transport in MOFs.
\end{abstract}
\maketitle

\section{Introduction}
In the last two decades, due to their ultra-high porosity \cite{rowsell2004metal} and structural tunability \cite{yaghi2003reticular}, \glspl{mof} have shown great potential in various applications, such as gas storage and separation \cite{li2014porous}, water harvesting \cite{hanikel2020mof}, electronic devices \cite{stassen2017updated}, and heterogeneous catalysis \cite{rogge2017metal}. \Gls{ltc} is a critical parameter for \glspl{mof} in the context of thermal energy conversion, thermal management and thermal stability and has attracted extensive experimental \cite{huang2007thermal2,ming2014thermophysical,erickson2015thin,semelsberger2016room,gunatilleke2017thermal,sun2017microporous,huang2019general,huang2020situ,babaei2020observation} and theoretical \cite{huang2007thermal1,zhang2013thermal,han2014relationship,babaei2016mechanisms,babaei2017effect,sezginel2018thermal,wieme2019thermal,wei2019impacts,sorensen2020metal,islamov2020influence,sezginel2020effect,babaei2020enhanced,ying2020impacts,ying2021effect,cheng2021molecular,zhang2021insights,lamaire2021atomistic,wieser2021identifying,zhou2021vibrational, zhou2022origin,wieser2022exploring,fan2022ultralong} studies. 

It is generally difficult to produce large-scale single crystals of \glspl{mof} which then usually exist in the form of powders. The small crystalline sizes present a great challenge for experimentally measuring the \gls{ltc} of \gls{mof} crystals \cite{huang2019general} because the contact thermal resistance between the crystalline particles can introduce large systematic errors. Despite this, single-crystalline MOF-5 \cite{li1999design} can be grown up to a linear dimension of a couple of millimeter and its \gls{ltc} has been measured to be \SI{0.32}{\watt\per\meter\per\kelvin} \cite{huang2007thermal2} using the steady-state direct method. Recent advances in experimental techniques also enabled the measurement of the \gls{ltc} of a few typical \glspl{mof} in single-crystal form. For example, the \gls{ltc} of single-crystal ZIF-8 \cite{huang2006ligand} was measured to be \SI{0.64 \pm 0.09}{\watt\per\meter\per\kelvin}, using the Raman-resistance temperature detectors method \cite{huang2020situ}. For HKUST-1 \cite{chui1999chemically}, a value of \SI{0.69 \pm 0.05}{\watt\per\meter\per\kelvin} was determined based on the thermoreflectance method \cite{babaei2020observation}.

Computationally, \gls{md} simulation has played an important role in calculating the \gls{ltc} and revealing the underlying mechanics for thermal transport in \glspl{mof}. Both the \gls{emd} \cite{green1954markoff, kubo1957statistical} and the \gls{nemd} methods have been extensively used. However, results from these two methods seem to be conflicting. For example, a weak temperature dependence of \gls{ltc} in MOF-5 has been predicted using \gls{emd} simulations, which was attributed to short phonon \glspl{mfp} on the order of the lattice parameter \cite{huang2007thermal1}. Similarly, for ZIF-4 and ZIF-62, it has been suggested, based on \gls{emd} simulations \cite{zhou2021vibrational, zhou2022origin}, that the phonon transport are strongly localized with phonon \glspl{mfp} less than \SI{1.5}{\nano\meter}. However, recent \gls{nemd} simulations suggested that the \gls{ltc} of MOF-5 is actually not fully converged even when the transport length exceeds \SI{20}{\nano\meter} \cite{wieser2021identifying, wieser2022exploring} and phonons with \glspl{mfp} larger than \SI{100}{\nano\meter} exist in HKUST-1 \cite{fan2022ultralong}. 

\begin{figure}[ht]
\centering
\includegraphics[width=\columnwidth]{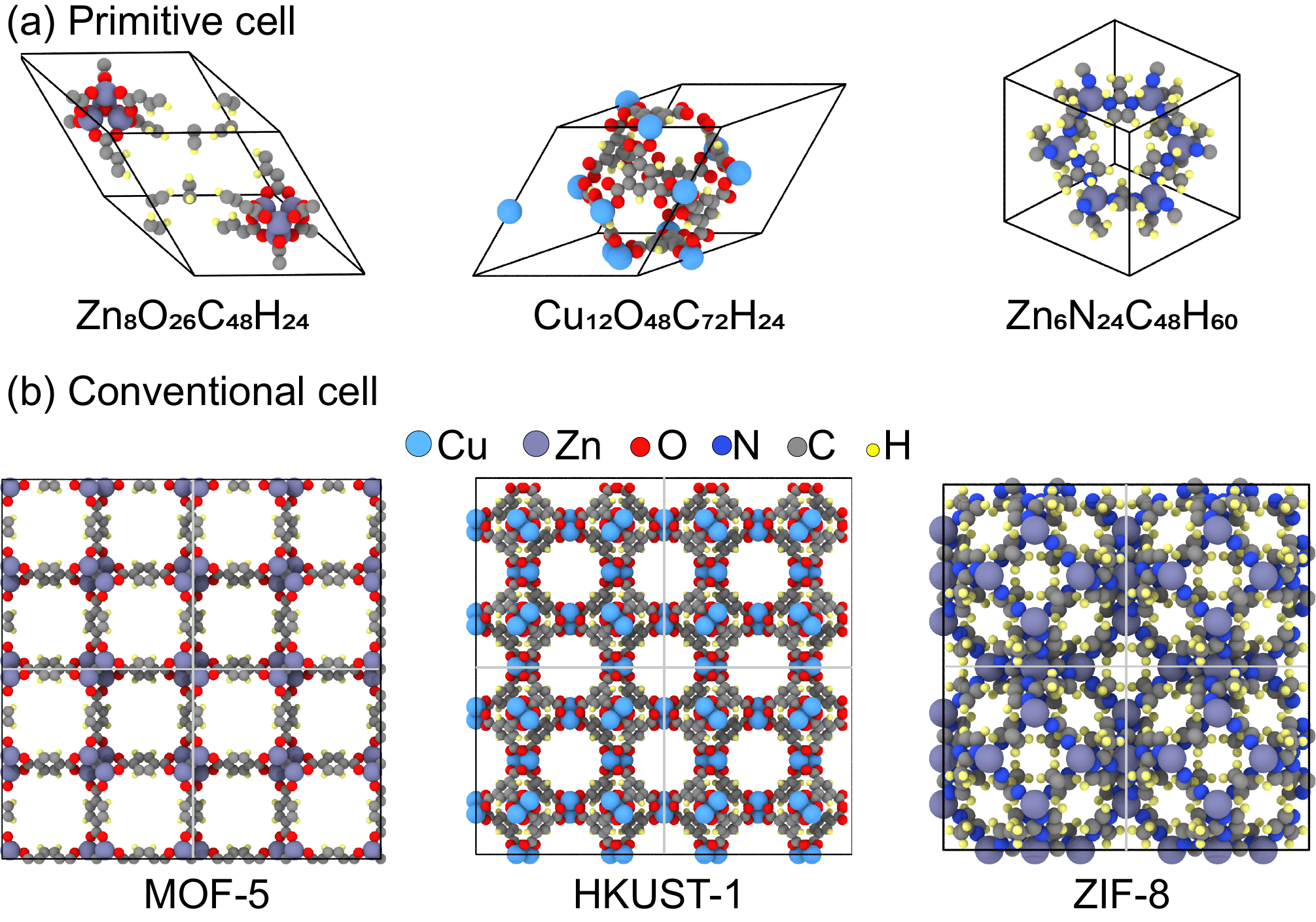}
\caption{The (a) primitive and (b) \numproduct{2x2x2} conventional cells of MOF-5 (left), HKUST-1 (middle), and ZIF-8 (right). The primitive cells contain 106, 156, and 138 atoms and the \numproduct{2x2x2} conventional ones contain 3992, 4992, and 2208 atoms. MOF-5 \cite{li1999design} is formed by connecting Zn$_4$O clusters with 1,4-benzenedicarboxylate organic linkers, resulting in a primitive cubic net topology. HKUST-1 \cite{chui1999chemically} features copper-paddle wheel units connected by 1,3,5-benzenetricarboxylate organic linkers, creating a body-centered cubic net topology. Lastly, ZIF-8 \cite{huang2006ligand} comprises of tetrahedral ZnN$_4$ nodes connected by 2-methylimidazolate organic linkers, generating a sodalite net topology. We use primitive cells to prepare the reference datasets for training. For the calculation of \gls{ltc} in \gls{md} simulations, we employ conventional cells of various system sizes. The \textsc{ovito} package \cite{stukowski2009visualization} was used for visualization.}
\label{figure:model}
\end{figure}

The differences in the results could be partially attributed to the different interatomic potentials used in different works, but there are two other possible causes. On the one hand, the heat current as implemented in the \textsc{lammps} package \cite{thompson2022lammps} used in most of the previous \gls{emd} simulations \cite{zhang2013thermal, han2014relationship, wei2019impacts, sorensen2020metal, islamov2020influence, ying2020impacts, ying2021effect, zhou2021vibrational, zhou2022origin} was incorrect for many-body potentials, which could lead to significantly reduced \gls{ltc} and phonon \glspl{mfp} in systems with low-dimensional features \cite{fan2015force}. On the other hand, it is often assumed that a supercell with \numproduct{2x2x2} conventional cells is sufficient to obtain convergent results in \gls{emd} simulations \cite{huang2007thermal1, zhang2013thermal, han2014relationship, wei2019impacts, sorensen2020metal, ying2020impacts, ying2021effect}, which is not necessarily valid. It is important to resolve the conflict between the \gls{emd} and the \gls{nemd} methods and obtain a clear and consistent picture of the phonon \glspl{mfp} in \glspl{mof} because phonon \glspl{mfp} determine the variation of the apparent \gls{ltc} with respect to the sample length, which, in the context of \glspl{mof}, can be the grain size of polycrystals or the linear size of the clusters in \gls{mof} powders.

In this paper, we present a coherent understanding of the phonon \glspl{mfp} in \glspl{mof} by considering three typical \glspl{mof}, including MOF-5 \cite{li1999design}, HKUST-1 \cite{chui1999chemically}, and ZIF-8 \cite{huang2006ligand}, as shown in \autoref{figure:model}. To this end, we construct accurate interatomic potentials for these materials by training state-of-the-art \glspl{mlp} against quantum-mechanical \gls{dft} calculations. Recently, \glspl{mlp} have been successfully applied to study a variety of properties of \glspl{mof} such as mechanical properties \cite{tayfuroglu2022neural}, phase transition \cite{vandenhaute2023machine}, and gas diffusion \cite{achar2022combined, zheng2023quantum}. Herein, we choose to use the \gls{nep} approach \cite{fan2021neuroevolution, fan2022improving, Fan2022GPUMD} that has been demonstrated to be highly efficient while being also sufficiently accurate. Moreover, the \gls{nep} approach has been implemented in the \textsc{gpumd} package \cite{Fan2017CPC}, which has the correct implementation of the heat current \cite{fan2015force}. We perform large-scale \gls{md} simulations to study thermal transport, establishing a coherent picture of thermal transport in \glspl{mof} by using three \gls{md} methods, including the aforementioned \gls{emd} and \gls{nemd} and a third one, namely, the \gls{hnemd} method \cite{Fan2019PRB}. Our results show that in all the three \glspl{mof}, the phonon \glspl{mfp} can reach the sub-micrometer scale in the low-frequency region, which indicates that the \gls{ltc} will be significantly reduced in polycrystals or powders with nanometer-scale grain sizes.

\section{Results and Discussion}

\subsection{Performance evaluation of the NEP models}

\begin{figure}[ht]
\center
\includegraphics[width=\columnwidth]{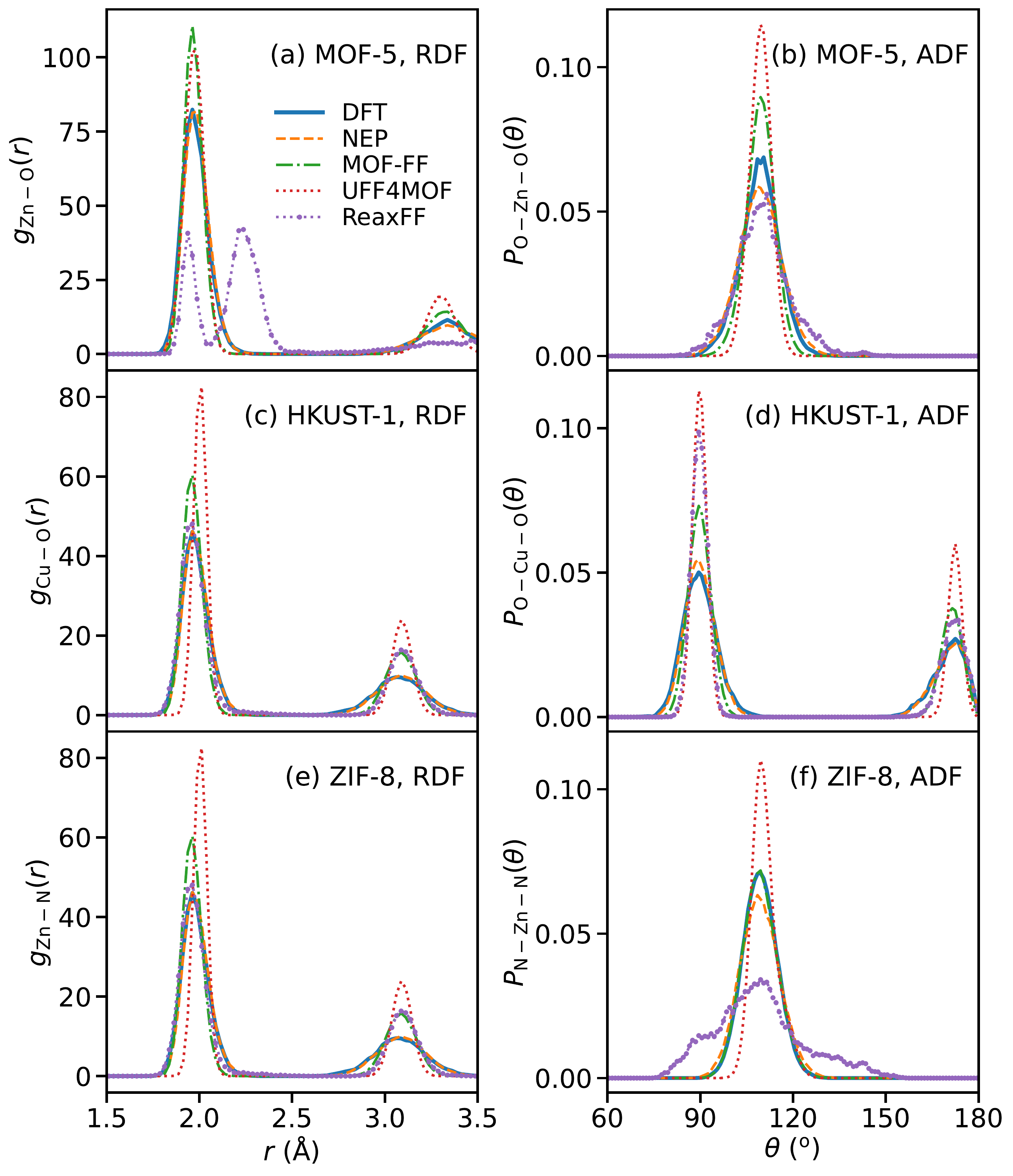}
\caption{Typical \glspl{rdf} (left) and \glspl{adf} (right) in MOF-5 (top), HKUST-1 (middle), and ZIF-8 (bottom) calculated using classical \gls{md} simulations at \SI{300}{\kelvin} driven by \gls{dft}, \gls{nep}, UFF4MOF \cite{addicoat2014extension,boyd2017force}, MOF-FF \cite{pallach2021frustrated, bureekaew2013mof, durholt2019ab}, and ReaxFF \cite{monti2013reactive, han2010molecular, yang2018enabling}.}
\label{figure:rdf}
\end{figure}

We employed an iterative approach to develop machine-learned \gls{nep} models, as illustrated in Fig. S1 (see \autoref{section:dataset} for details). The \gls{nep} models for all the three \glspl{mof} achieve very high accuracy, with \glspl{rmse} of energy, force, and virial being less than \SI{0.6}{\milli\electronvolt\per\atom}, \SI{60}{\milli\electronvolt\per\angstrom}, and \SI{5}{\milli\electronvolt\per\atom}, respectively, for both the training and test data sets (see Fig. S2 and Table S1 for details). The zero-temperature lattice constants predicted by the \gls{nep} models closely match those predicted by \gls{dft} calculations, with a relatively error being less than 0.1 \% (see Table S2 for details).

To further evaluate the accuracy of the \gls{nep} models in \gls{md} simulations, we compare the \glspl{rdf} and \glspl{adf} of the three \glspl{mof} at \SI{300}{\kelvin} calculated from the \gls{nep} models with those obtained from \gls{dft} and a few representative empirical force fields (see \autoref{figure:rdf}). The selected force fields include MOF-FF \cite{bureekaew2013mof}, UFF4MOF \cite{addicoat2014extension}, and ReaxFF \cite{van2001reaxff}, which have been widely used in \gls{md} simulation of the thermodynamics of MOF-5 \cite{pallach2021frustrated,banlusan2015mechanisms,islamov2023high}, HKUST-1 \cite{evans2019assessing,wang2023effects,islamov2023high}, and ZIF-8 \cite{ying2020impacts, sorensen2020metal, islamov2023high}. 

\begin{figure}[ht]
\center
\includegraphics[width=\columnwidth]{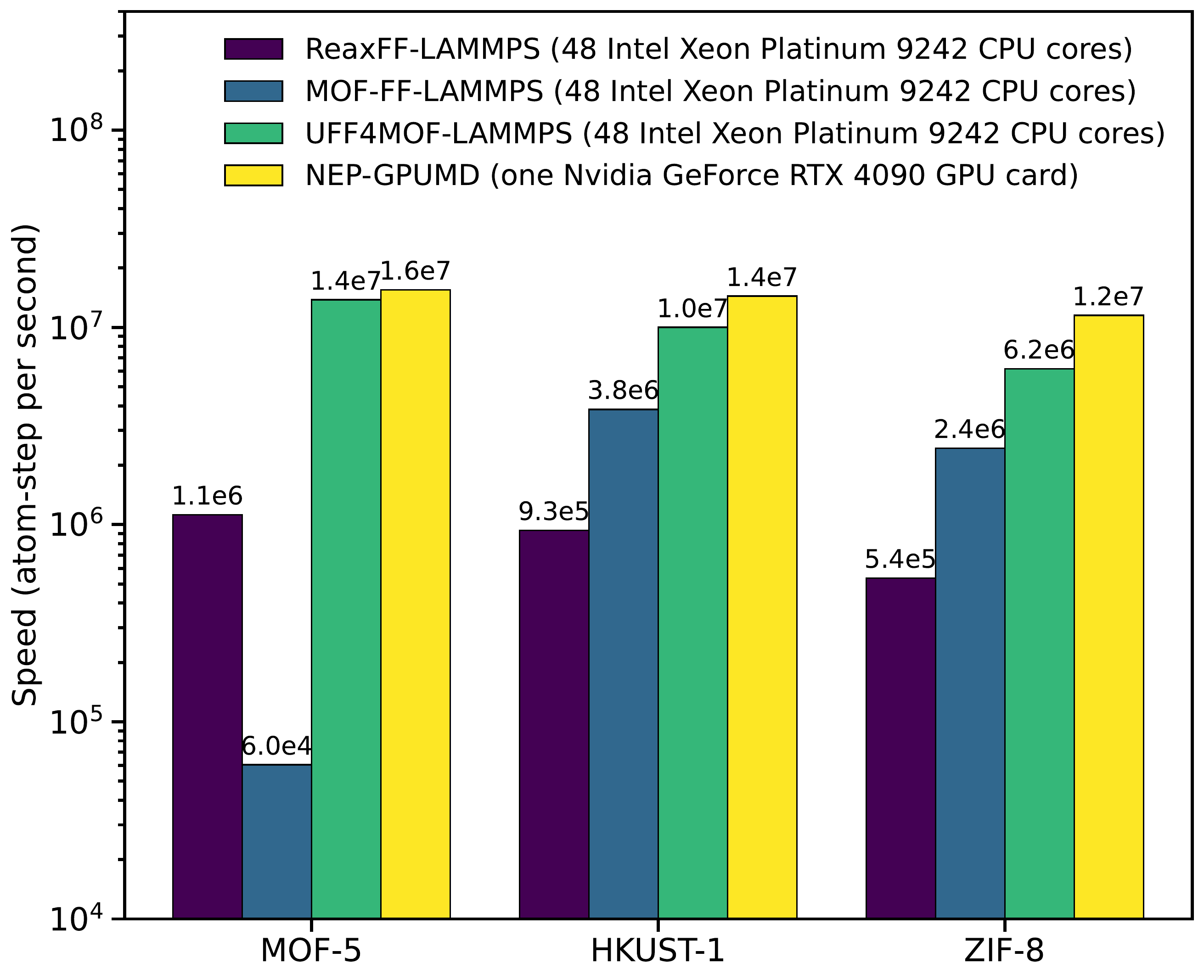}
\caption{Computational speeds of the \gls{nep} models as implemented in \textsc{gpumd} \cite{Fan2017CPC} (version 3.7, running with an Nvidia GeForce RTX 4090 GPU card) and several empirical force fields as implemented in \textsc{lammps} \cite{thompson2022lammps} (version 28 Mar 2023, running with 48 Intel Xeon Platinum 9242 CPU cores) for the three \glspl{mof}. The benchmark was performed using a \numproduct{5x5x5} supercell  containing \num{53000}, \num{78000}, and \num{34500} atoms for MOF-5, HKUST-1, and ZIF-8, respectively.}
\label{figure:speed}
\end{figure}

For \gls{rdf}, we considered the most important pairs at the metal-linker interfaces: Zn-O in MOF-5, Cu-O in HKUST-1, and Zn-N in ZIF-8. Accordingly, we considered the O-Zn-O, O-Cu-O, and N-Zn-N triplets for the \gls{adf}. For both \gls{rdf} and \gls{adf}, \gls{nep} achieves excellent agreement with \gls{dft}, while the empirical force fields exhibit varying degrees of discrepancies compared to \gls{dft}. Among the three selected empirical force fields, MOF-FF achieves the highest accuracy and the other two are much less accurate. UFF4MOF predicts inaccurate positions for some \gls{rdf} peaks and too sharp peaks in the \glspl{adf}, while ReaxFF predicts a non-existing peak in the \gls{rdf} of MOF-5 and too flat \gls{adf} of ZIF-8. The inaccuracies of  ReaxFF in \gls{md} simulations of ZIFs have also been highlighted elsewhere \cite{castel2022atomistic}.

Our \gls{nep} models not only achieve high accuracy, but also have high computational efficiency. As can be seen from \autoref{figure:speed}, the \gls{nep} models as implemented in \textsc{gpumd} \cite{Fan2017CPC} running with a single Nvidia GeForce RTX 4090 GPU card are faster than all the empirical force fields as implemented in \textsc{lammps} \cite{thompson2022lammps} running with 48 Intel Xeon Platinum 9242 CPU cores. The MOF-FF force field for MOF-5 \cite{pallach2021frustrated} is particularly slow because it evaluates the Coulomb potential using the expensive Ewald summation. Apart from this, the computational speed of MOF-FF is higher than that of ReaxFF and lower than that of UFF4MOF.

\begin{figure}[ht]
\includegraphics[width=\columnwidth]{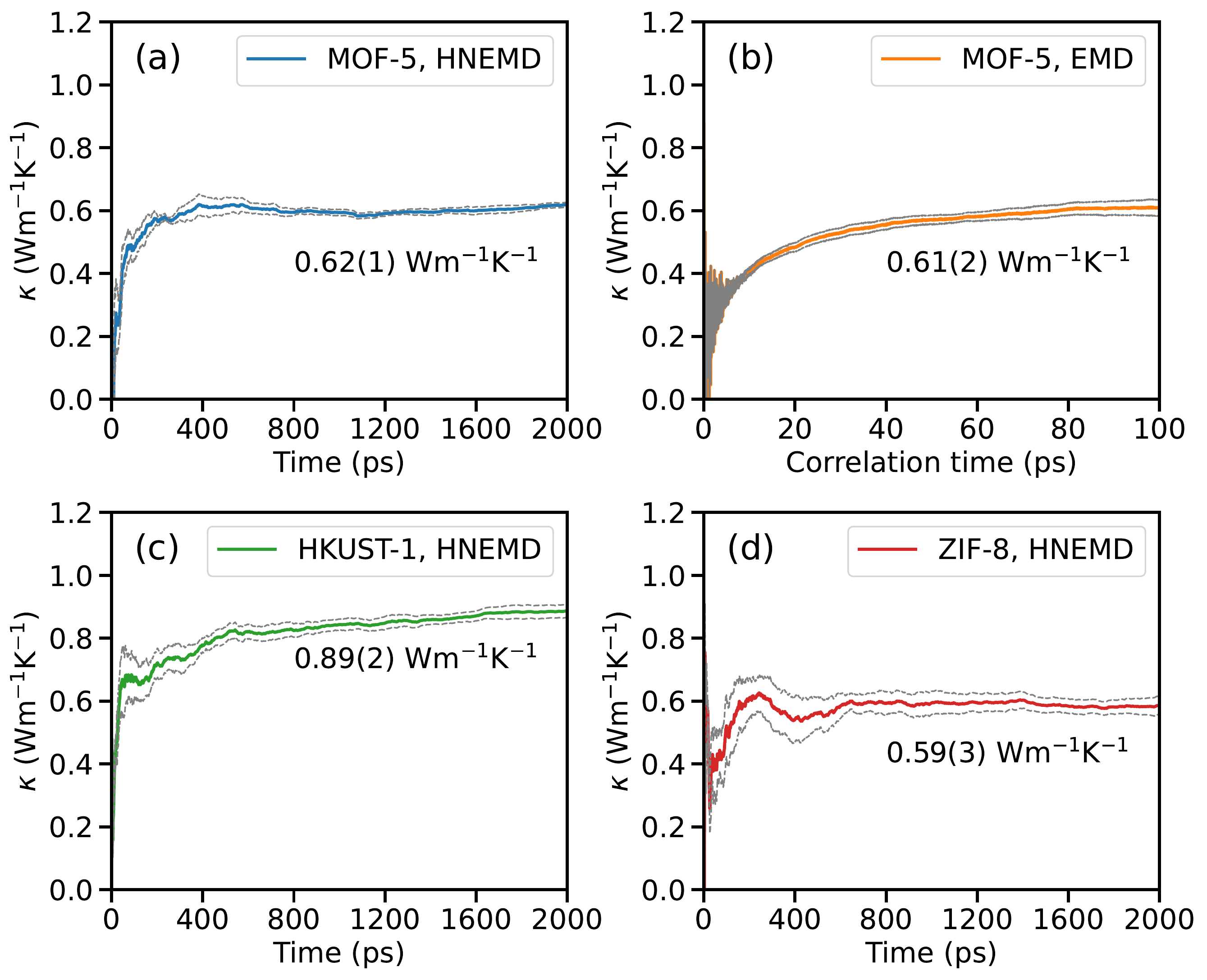}
\caption{Cumulative average of the \gls{ltc} as calculated using the \gls{hnemd} method for (a) MOF-5, (c) HKUST-1, and (d) ZIF-8 at \SI{300}{\kelvin}. (b) Running \gls{ltc} as calculated using the \gls{emd} method for MOF-5. The simulation cells contain \numproduct{5x5x5} conventional cells. In each panel, the middle line represents the average and the other two represent the upper and lower bounds as calculated from five independent runs. }
\label{figure:kappa}
\end{figure}

\subsection{Thermal transport in MOFs}

After confirming the reliability of our \gls{nep} models in \gls{md} simulations, we apply them to calculate the \gls{ltc} using the various \gls{md} methods as reviewed in \autoref{section:methods}. In \autoref{figure:kappa} we show the \gls{hnemd} and \gls{emd} results for the three \glspl{mof} at 300 K and zero pressure obtained by using a supercell with \numproduct{5x5x5} conventional cells. In both methods, the \gls{ltc} has converged with respect to the production or correlation time. The results from the two methods are consistent as expected based on their physical equivalence \cite{Fan2019PRB}. However, we note that even if the total production time we used for the \gls{emd} method for each system (\SI{50}{\nano\second}) is much larger than that for the \gls{hnemd} method (\SI{10}{\nano\second}), the latter still has smaller statistical errors, which demonstrates the superior computational efficiency of the \gls{hnemd} method \cite{Fan2019PRB}.

\begin{figure}[ht]
\center
\includegraphics[width=\columnwidth]{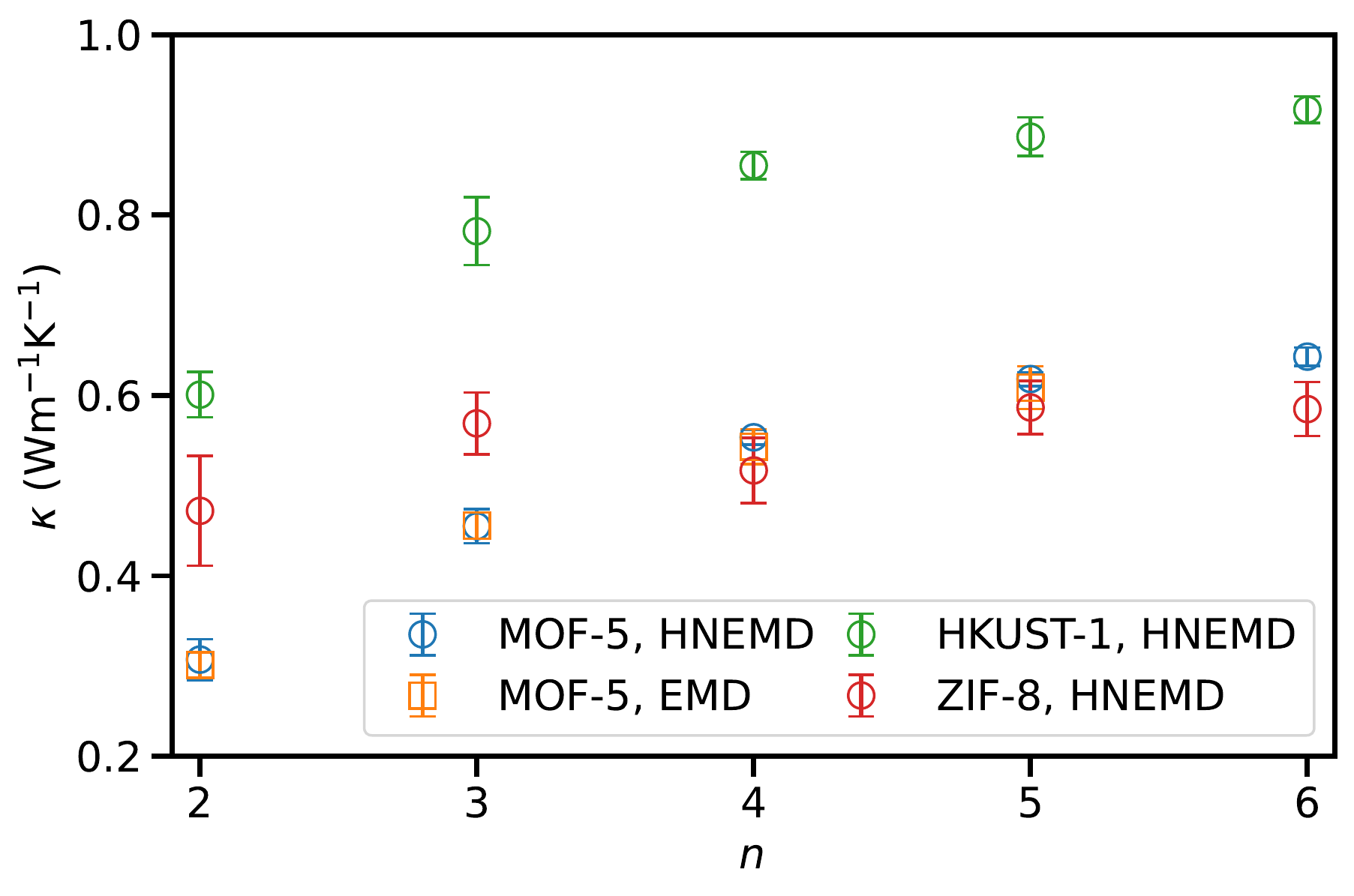}
\caption{\gls{ltc} $\kappa$ as a function of the number of conventional cells $n$ in one direction of a cubic supercell for  the three \glspl{mof} at \SI{300}{\kelvin} from the \gls{hnemd} and \gls{emd} (for MOF-5 only) simulations.}
\label{figure:size1}
\end{figure}

In both the \gls{hnemd} and the \gls{emd} methods, periodic boundary conditions are applied in the three directions and the results are regarded as those for infinitely large systems. However, this requires to eliminate the finite-size effects by using a sufficiently large supercell. While previous works often assumed that a supercell with \numproduct{2x2x2} conventional cells is sufficient to obtain convergent results in \gls{emd} simulations \cite{huang2007thermal1, zhang2013thermal, han2014relationship, wei2019impacts, sorensen2020metal, ying2020impacts, ying2021effect}, our results in \autoref{figure:size1} show that for all the three \glspl{mof}, a supercell with \numproduct{5x5x5} conventional cells is required to converge the \gls{ltc}. Note that agreement between the \gls{hnemd} and the \gls{emd} methods for MOF-5 are confirmed for all the simulation domains up to the \numproduct{5x5x5} supercell. 

\begin{table}[thb]
\setlength{\tabcolsep}{0.7Mm}
\caption{\glspl{ltc} (in units of $\rm{W/(m\ K)}$) of MOF-5 calculated by using the \gls{hnemd} method with different supercell sizes at different temperatures. The numbers within the parentheses are statistical uncertainties for the last significant number.}
\begin{tabular}{llllll}
\hline
\hline
Supercells        & 200 K & 250 K & 300 K & 350 K & 400 K \\
\hline
\numproduct{2x2x2} & 0.38(4) & 0.36(4) & 0.31(2) & 0.31(2) & 0.31(2) \\
\numproduct{5x5x5} & 0.76(1) & 0.72(2) & 0.62(1) & 0.55(1) & 0.51(2) \\
\hline
\hline
\end{tabular}
\label{table:temp}
\end{table}

\begin{figure}[ht]
\center
\includegraphics[width=\columnwidth]{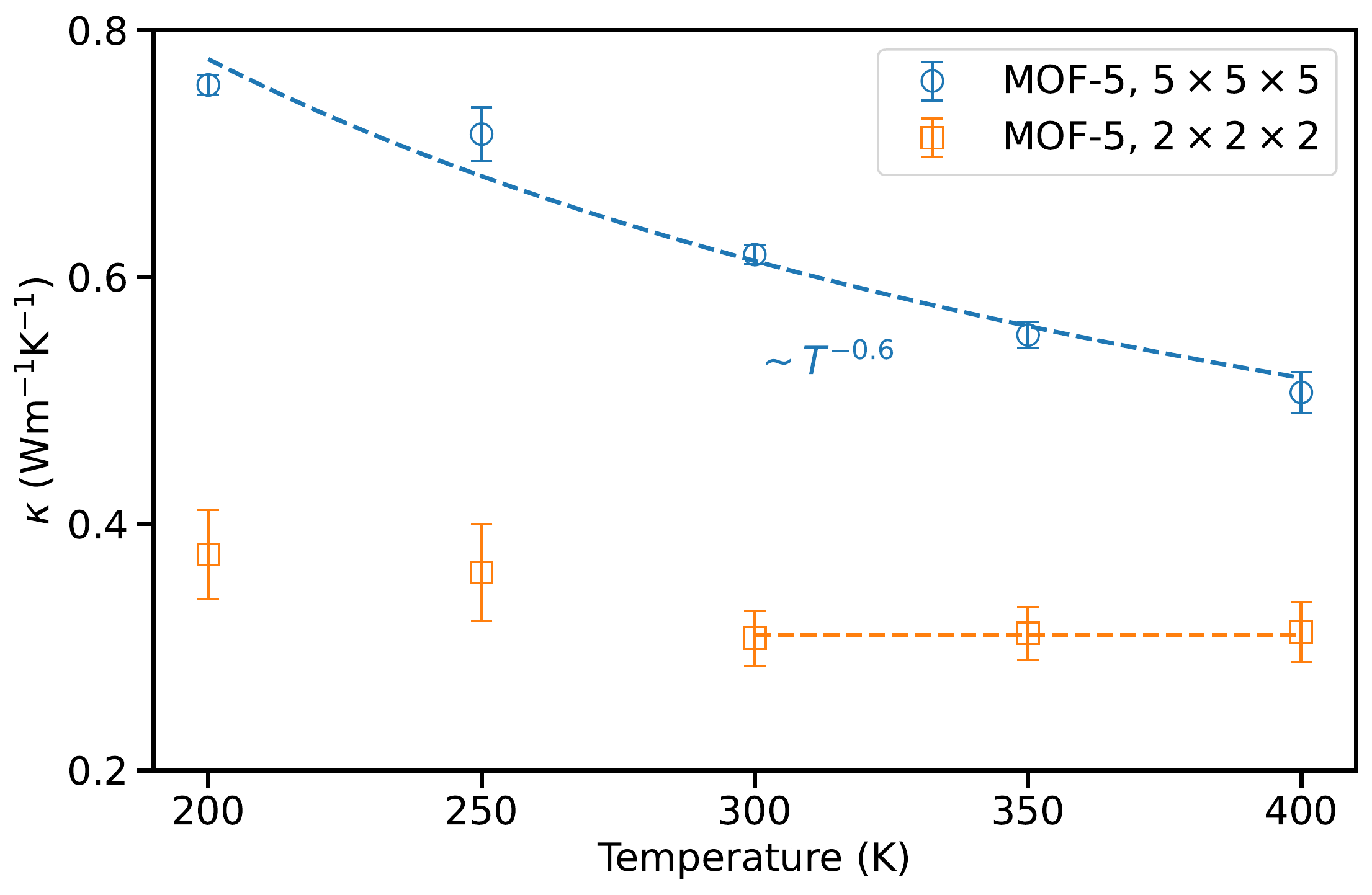}
\caption{\gls{ltc} $\kappa$ as a function of temperature for MOF-5 as obtained by using the \numproduct{5x5x5} and \numproduct{2x2x2} supercells. The dashed line for the case of \numproduct{5x5x5} supercell indicates a $\sim T^{-0.6}$ fitting, and the dashed line for the case of \numproduct{2x2x2} supercell indicates that the \gls{ltc} is a constant from 300 to 400 K.}
\label{figure:temp}
\end{figure}

Several studies have predicted a weak temperature dependence of the \gls{ltc} of MOF-5 \cite{huang2007thermal1} and various ZIF materials, such as ZIF-8 \cite{zhang2013thermal, ying2020impacts}, ZIF-4 \cite{zhou2022origin}, and ZIF-62 \cite{zhou2022origin}, using a small supercell in \gls{emd} simulations. With the small \numproduct{2x2x2} supercell, we also obtained a weak temperature dependence of the \gls{ltc} of MOF-5 as shown in \autoref{figure:temp} and \autoref{table:temp}. Particularly, the \gls{ltc} is almost a constant from 300 to \SI{400}{\kelvin}. However, using the converged \numproduct{5x5x5} supercell, a notable temperature dependence of $\sim T^{-0.6}$ is obtained. This again highlights the importance of using a sufficiently large supercell in \gls{emd} and \gls{hnemd} simulations. 

\begin{table}[thb]
\setlength{\tabcolsep}{2.5Mm}
\caption{\glspl{ltc} (in units of $\rm{W/(m\ K)}$) of the three \glspl{mof} calculated by using the \gls{hnemd} method and \gls{emd} method (for MOF-5 only) with different supercell sizes (the number of conventional cells $n$ in one direction of a cubic supercell) at \SI{300}{\kelvin}. The numbers within the parentheses are statistical uncertainties for the last significant number.}
\label{table:data}
\begin{tabular}{lllll}
\hline
\hline
& \multicolumn{1}{c}{EMD} & \multicolumn{3}{c}{HNEMD}\\
$n$ & MOF-5 & MOF-5 & HKUST-1 & ZIF-8 \\
\hline
2   & 0.30(1) & 0.31(2) & 0.60(3) & 0.47(6) \\
3   & 0.46(2) & 0.46(2) & 0.78(4) & 0.57(3) \\
4   & 0.54(2) & 0.55(1) & 0.86(2) & 0.52(4) \\
5   & 0.61(2) & 0.62(1) & 0.89(2) & 0.59(3) \\
6   & /       & 0.64(1) & 0.92(2) & 0.59(3) \\
\hline
\hline
\end{tabular}
\label{table:size}
\end{table}

The calculated \gls{ltc} values at 300 K using the converged \numproduct{5x5x5} supercell are listed in \autoref{table:size}. Our predicted \gls{ltc} of ZIF-8, \SI{0.59(3)}{\watt\per\meter\per\kelvin}, closely matches with the experimentally measured value of \SI{0.64(9)}{\watt\per\meter\per\kelvin} \cite{huang2020situ}. For HKUST-1, our predicted \gls{ltc}, \SI{0.89(2)}{\watt\per\meter\per\kelvin} is only slightly larger than the experimentally measured value of \SI{0.69(5)}{\watt\per\meter\per\kelvin} \cite{babaei2020observation}. For these two \glspl{mof}, the good agreement with experiments is achieved for the first time.
In the case of MOF-5, we predicted a \gls{ltc} of \SI{0.62(1)}{\watt\per\meter\per\kelvin} that is significantly higher than the experimentally measured value of \SI{0.32}{\watt\per\meter\per\kelvin} \cite{huang2007thermal2}. This discrepancy is yet to be understood and we will discuss it further below.

\begin{figure}[ht]
\begin{center}
\includegraphics[width=\columnwidth]{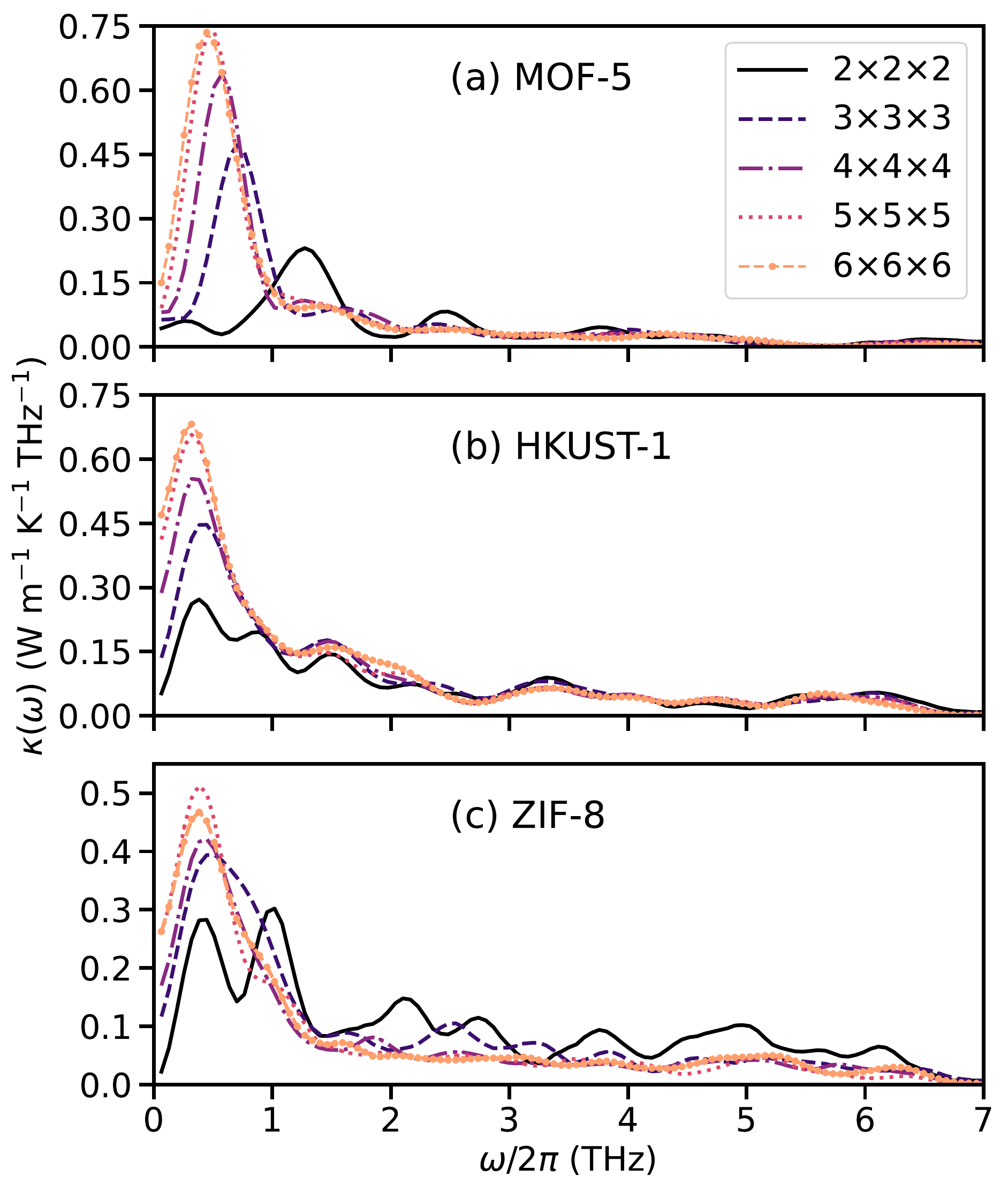}
\caption{Spectral \gls{ltc} $\kappa(\omega)$ as a function of phonon frequency $\omega/2\pi$ for (a) MOF-5, (b) HKUST-1, and (c) ZIF-8 at \SI{300}{\kelvin} calculated by using different supercell sizes.}
\label{figure:size2}
\end{center}
\end{figure}

\begin{figure}[ht]
\begin{center}
\includegraphics[width=\columnwidth]{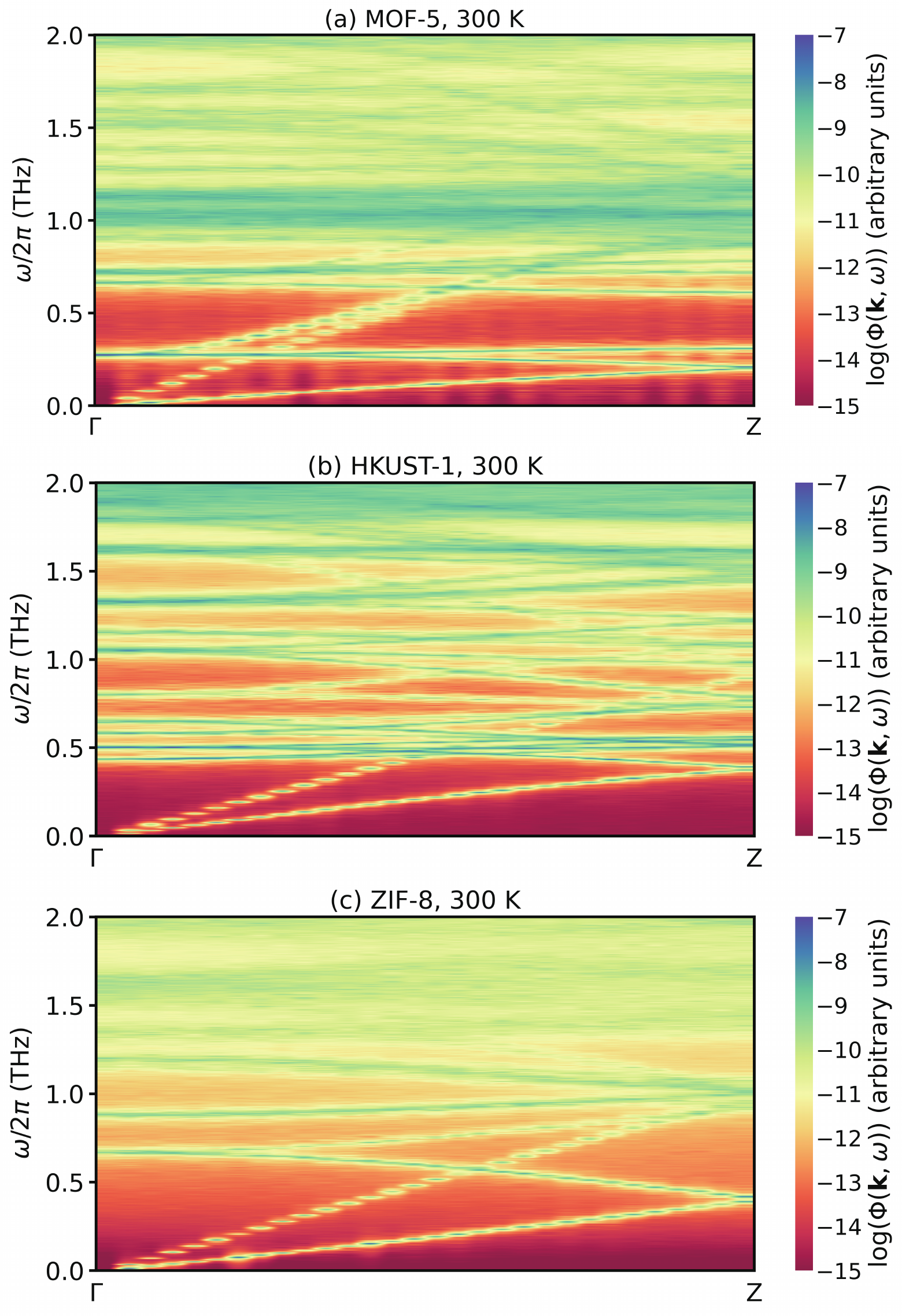}
\caption{Phonon \gls{sed} of (a) MOF-5, (b) HKUST-1, and (c) ZIF-8 at \SI{300}{\kelvin} as a function of wave number in the $\Gamma$-Z path and phonon frequency.}
\label{figure:sed}
\end{center}
\end{figure}

The finite-size effects can be understood in terms of the spectral \gls{ltc} as shown in \autoref{figure:size2}. With a larger simulation domain size, phonons with longer wave lengths (smaller frequencies) are brought into existence which contribute significantly in the low-frequency region. \autoref{figure:size2} shows that the contribution from phonons with $\omega/2\pi<1$ THz is significantly reduced when the small \numproduct{2x2x2} supercell is used. To understand this more quantitatively, we note that the phonon group velocities for the acoustic branches are about \SI{5}{\kilo\meter\per\second}, based on the \gls{sed} shown in \autoref{figure:sed}. Taking MOF-5 as an example, the largest phonon wave length that can exist in a \numproduct{2x2x2} supercell is about \SI{5}{\nano\meter} and the frequency for it is about \SI{1}{\tera\hertz}. That is, phonons with frequencies smaller than \SI{1}{\tera\hertz} are largely suppressed in this supercell. With a \numproduct{5x5x5} supercell, phonons with $\omega/2\pi<1$ THz can be activated and their contribution to the total \gls{ltc} is almost converged. In all the three \glspl{mof}, phonons with $\omega/2\pi<1$ THz contribute about 50\% to the total \gls{ltc}.

\begin{figure}[ht]
\includegraphics[width=\columnwidth]{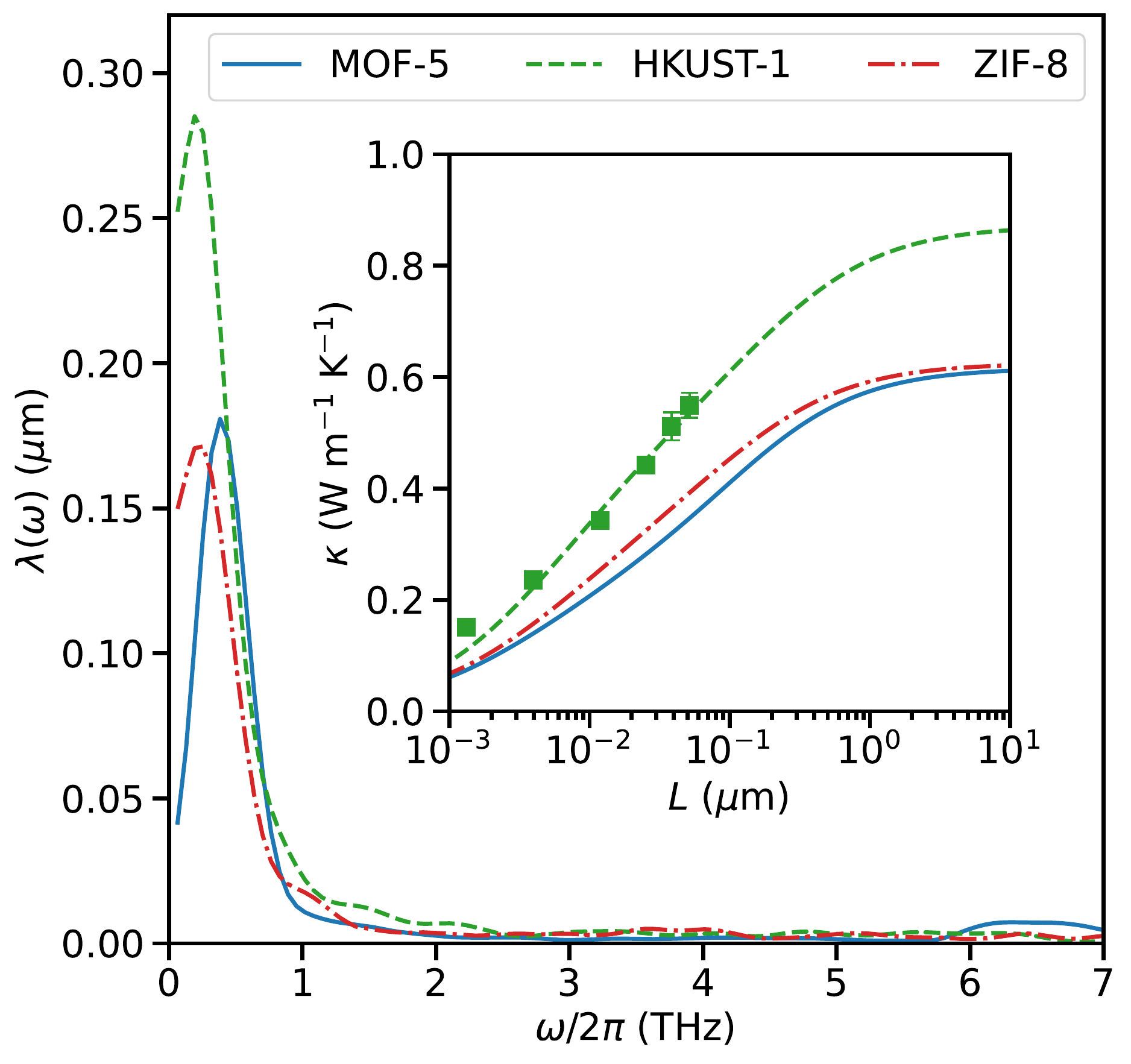}
\caption{Phonon \gls{mfp} $\lambda(\omega)$ as a function of phonon frequency $\omega/2\pi$ for the three \glspl{mof} at \SI{300}{\kelvin}. Inset: Apparent \gls{ltc} $\kappa(L)$ as a function of the transport length $L$ for the three MOFs at \SI{300}{\kelvin} as calculated from the \gls{hnemd}-based spectral decomposition method (lines) and direct \gls{nemd} simulations (markers; for HKUST-1 only).}
\label{figure:mfp}
\end{figure}

The large spectral \glspl{ltc} in the low-frequency region are associated with the large phonon \glspl{mfp}, as shown in \autoref{figure:mfp}. The maximum phonon \glspl{mfp} for MOF-5, HKUST-1, and ZIF-8 are \SI{0.17}{\micro\meter}, \SI{0.28}{\micro\meter}, and \SI{0.16}{\micro\meter}, respectively. Due to the large \glspl{mfp}, the apparent \gls{ltc} only saturates in the micrometre scale of system length $L$, as shown in the inset of \autoref{figure:mfp}. The slow convergence of the \gls{ltc} with respect to $L$ for HKUST-1 is further confirmed by independent \gls{nemd} simulations.

Using an empirical force field, Wieser \textit{et al.}  \cite{wieser2021identifying, wieser2022exploring} observed that the \gls{ltc} of MOF-5 is not converged up to \SI{20}{\nano\meter} in \gls{nemd} simulations, which is consistent to our results. This means that the large phonon \glspl{mfp} in \glspl{mof} are not sensitive to the details of the interatomic potential. As we have discussed above, the large phonon \glspl{mfp} in the low-frequency region means that a \numproduct{2x2x2} supercell is not sufficient to obtain converged \gls{ltc} in \gls{emd} simulations. With this small supercell, our \gls{emd} prediction of \SI{0.30(1)}{\watt\per\meter\per\kelvin} is very close to the experimental value \cite{huang2007thermal2} of \SI{0.32}{\watt\per\meter\per\kelvin} as well as the previous \gls{emd} simulations \cite{huang2007thermal1}. However, this good agreement is most likely an accident as the \gls{ltc} predicted by our \gls{emd} simulations with a \numproduct{5x5x5} supercell is \SI{0.61(2)}{\watt\per\meter\per\kelvin}, which is about twice of the experimental value. A possible explanation of the discrepancy between our predictions and the previous experiment \cite{huang2007thermal2} is that the measured samples might not be single crystals. According to the length-scaling in the inset of \autoref{figure:mfp}, the \gls{ltc} of MOF-5 is only about half of the converged value when $L$ is about \SI{0.1}{\micro\meter}. We therefore speculate that the experimental samples \cite{huang2007thermal2} might be nano-crystals with a characteristic size of \SI{0.1}{\micro\meter} or contain defects that have been observed to significantly reduce the thermal conductivity of \glspl{mof} \cite{islamov2020influence}.

\section{Summary and conclusions \label{section:summary}}

In summary, we have developed a set of accurate and highly efficient \glspl{mlp} for three typical \glspl{mof}, including MOF-5, HKUST-1, and ZIF-8, using the efficient \gls{nep} approach and quantum-mechanical \gls{dft} calculations. For each \gls{mof}, the \gls{nep} model accurately reproduces the \gls{rdf} and \gls{adf} as compared to \gls{dft} calculations, showing much higher accuracy than typical empirical force fields. We have performed extensive \gls{md} simulations with the \gls{nep} models to study thermal transport in these \glspl{mof}. We have mainly used the \gls{hnemd} method, but have also used the \gls{emd} and \gls{nemd} methods for crosschecking the consistency of the results. 

We have carefully examined the finite-size effects in \gls{emd} and \gls{hnemd} simulations for which previous works have not paid sufficient attentions. Contrary to the common assumption that a supercell with \numproduct{2x2x2} conventional cells is sufficient, our results indicate that at least a \numproduct{5x5x5} supercell is required to achieve converged results. This is not only relevant to the calculated \gls{ltc} values at a specific temperature, but also crucial for obtaining the correct temperature dependence of the \gls{ltc}. Our results suggest that the \glspl{ltc} of \glspl{mof} exhibit a moderate temperature dependence of $\sim T^{-0.6}$, which is not as strong as $\sim T^{-1}$ in conventional crystals, but is also not as weak as $\sim T^{0}$ in typical amorphous materials.

With the converged simulation cell size, we have obtained the spectral \gls{ltc} within the \gls{hnemd} method and found that phonons with frequency $\omega/2\pi<1$ THz contribute about half to the total \gls{ltc}. These phonons have sub-micrometer \glspl{mfp}, which can neither be accessed by \gls{emd} simulations with a small supercell nor \gls{nemd} simulations with short domain lengths, making them largely undiscovered before. 

Among the three \glspl{mof} we considered, HKUST-1 has the largest \gls{ltc}, while the other two have smaller and comparable \gls{ltc}. Our results for HKUST-1 and ZIF-8 are in very good agreement with experiments \cite{huang2020situ,babaei2020observation}, but the same has not been achieved for MOF-5 \cite{huang2007thermal2}. More work is needed to resolve this discrepancy. 

Overall, we have obtained a clear picture of thermal transport in \glspl{mof}. They typically have a \gls{ltc} smaller than \SI{1}{\watt\per\meter\per\kelvin} but there are THz-phonons with sub-micrometer \glspl{mfp} that contribute a significant portion of the \gls{ltc} and give rise to a strong length dependence of the apparent \gls{ltc}. Therefore, the \gls{ltc} in nano-crystals can be heavily reduced compared to single-crystals.

\section*{Notes}

The authors declare no competing financial interest.
The source code and documentation for \textsc{gpumd} are available
at \url{https://github.com/brucefan1983/GPUMD} and \url{https://gpumd.org}, respectively. The source code for \gls{sed} calculations is avaliable at \url{https://github.com/Tingliangstu/pySED}. The training and testing results for the NEP models are freely available at \url{https://gitlab.com/brucefan1983/nep-data}. Representative input and output files for thermal conductivity calculations are freely available at \url{https://github.com/hityingph/supporting-info}.

\section*{Acknowledgments}

The authors would like to thank Bing Wang, Yanzhou Wang, Zezhu Zeng, and Shiyun Xiong for insightful discussions. 
P.Y. and Z.Z. acknowledge the National Natural Science Foundation of China (Grant No. 11932005) and
the program of Innovation Team in Universities and Colleges in Guangdong (2021KCXTD006). 
T.L. and J.X. acknowledge the Research Grants Council of Hong Kong (Grant No. AoE/P-701/20). 
J.Z. acknowledges the support from the Guangdong Basic and Applied Basic Research Foundation (No.2022A1515010631). Z.F. acknowledges support from the National Natural Science Foundation of China (No. 11974059). 

\appendix

\section*{Methods}
\label{section:methods}

\subsection{The \texorpdfstring{\gls{nep}}{NEP} approach}

In the \gls{nep} approach, a feedforward \gls{nn} with a single hidden layer of $N_\mathrm{neu}$ neurons is used to represent the site energy $U_{i}$ of atom $i$ as a function of a descriptor vector with $N_\mathrm{des}$ components \cite{fan2021neuroevolution}:
\begin{equation}
\label{equation:Ui}
U_i = \sum_{\mu=1}^{N_\mathrm{neu}}w^{(1)}_{\mu}\tanh\left(\sum_{\nu=1}^{N_\mathrm{des}} w^{(0)}_{\mu\nu} q^i_{\nu} - b^{(0)}_{\mu}\right) - b^{(1)},
\end{equation}
where $\tanh(x)$ is the activation function, $\mathbf{w}^{(0)}$, $\mathbf{w}^{(1)}$, $\mathbf{b}^{(0)}$, and $b^{(1)}$ are the trainable weight and bias parameters in the \gls{nn}. The atomic environment descriptor $q^i_{\nu}$ consists of a number of radial and angular components similar to the Behler-Parrinello approach \cite{behler2007prl}. The radial descriptor components (with a cutoff radius $r_{\rm c}^{\rm R}$) only depend on atom-pair distances and are constructed based on Chebyshev polynomials. The angular descriptor components (with a cutoff radius $r_{\rm c}^{\rm A}$) also depend on angular information and are constructed based on spherical harmonics similar to the atomic cluster expansion approach \cite{drautz2019atomic}. A \gls{nep} model is trained using the \gls{snes} \cite{Schaul2011} by minimizing a loss function that is defined as a weighted sum of the \glspl{rmse} of energy, force, and virial as well as terms serving as $\ell_1$ and $\ell_2$ regularization. For more details of the \gls{nep} approach, we refer to the literature.~\cite{fan2021neuroevolution,fan2022improving,Fan2022GPUMD}.

There are a number of important hyperparameters in the \gls{nep} approach that need to be carefully chosen. Based on previous publications \cite{fan2021neuroevolution, fan2022improving, Fan2022GPUMD, wang2022quantum, ying2023variable, ying2023atomistic, dong2022anisotropic} and our extensive tests, we reached a set of optimized hyperparameters for the three \glspl{mof}. For all the three \glspl{mof}, we used the same hyperparameters except for the cutoff radius of the angular descriptor components $r_{\rm c}^{\rm A}$ that was set to \SI{4}{\angstrom} for both MOF-5 and HKUST-1, but to \SI{3}{\angstrom} for ZIF-8. The cutoff radius of the radial descriptor components was set to a larger value of \SI{6}{\angstrom} to incorporate more neighbors. Both the radial and angular descriptor components have radial functions. For the radial descriptor components, we used 13 radial functions, each being a linear combination of 13 Chebyshev polynomials. For the angular descriptor components, we used 9 radial functions, each being a linear combination of 9 Chebyshev polynomials. For the angular descriptor components, we used both three-body and four-body correlations up to degree $l=4$ and $l=2$ respectively in the spherical harmonics $Y_{lm}$. The number of neurons in the hidden layer was chosen to be $N_{\rm neu}=80$, which is large enough to achieve a high accuracy. 

\subsection{Reference data generation}
\label{section:dataset}
We employed an iterative scheme as shown in Fig. S1 to construct the training data for the \gls{nep} model of each material. While a more sophisticated method is needed for complex materials with very large primitive cells \cite{eckhoff2019molecular}, it is feasible to generate the training data using the primitive cells in our case. Starting from the optimized structure of the primitive cell of a given material, we performed constant-volume \gls{dft}-driven \gls{md} simulations with the target temperature linearly increasing from \SI{10}{\kelvin} to \SI{800}{\kelvin} within \SI{10}{\pico\second} (10000 steps) and sampled 560 structures. We also generated 100 structures by applying random cell deformations (-3\% to 3\%) and atomic displacements (within \SI{0.1}{\angstrom}) staring from the optimized structure. We performed more accurate \gls{dft} calculations (see below) for the 660 structures and obtained the initial data set and trained the first \gls{nep} model. Then we performed \gls{nep}-driven \gls{md} simulations at our intended thermodynamic conditions and selected 100 more structures based on the farthest-point sampling by comparing the distance of structures in \gls{2d} \gls{pc} space. We then trained the second \gls{nep} model based on the expanded 760 structures. Another round of expansion was similarly performed and our final data set has 860 structures for each material. We have used 90\% for training and 10\% for testing. 

The \gls{dft}-\gls{md} simulations were conducted in the isothermal ($NVT$) ensemble with an energy threshold of \SI{e-5}{\eV}, an energy cutoff of \SI{520}{\eV} for the electronic self-consistent loop, and the $\Gamma$ point was sampled in the Brillouin zone. To obtain the energy, force, and virial data for \gls{nep} training, we performed \gls{dft} calculations using the Perdew-Burke-Ernzerhof functional \cite{Perdew1996PRL} and the projector-augmented wave method \cite{Blchl1994PRB} implemented in Vienna ab-initio simulation package package \cite{Kresse1996PRB, Kresse1999PRB}. We sampled the Brillouin zone with a $k$-point density of \SI{0.2}{\per\angstrom} and used Gaussian smearing with a width of \SI{0.05} {\eV}. For the electronic self-consistent loop, we set a threshold of \SI{e-7}{\eV} with an energy cutoff of \SI{600}{\eV}. 

\subsection{The \texorpdfstring{\gls{hnemd}}{HNEMD} method and spectral conductivity in the diffusive regime}

The \gls{hnemd} method has been shown to be one of the most efficient \gls{md} methods for obtaining diffusive thermal transport properties \cite{Fan2019PRB,ying2023variable,dong2022anisotropic,ying2022thermal}.
In \gls{hnemd} simulation of thermal conduction in solid, an external driving force 
\begin{equation}
\label{equation:Fe}
\bm{F}_{i}^{\rm ext} = \bm{F}_{\rm e}\cdot\textbf{W}_{i}.
\end{equation}
is added to each atom $i$ to drive the system out of equilibrium. Here $\textbf{W}_{i}$ is the virial tensor of atom $i$ and $\bm{F}_{\rm e}$ is the driving force parameter which is of the dimension of inverse length. The driving force will induce a net heat current $\bm{J}$ whose ensemble average (represented by $\langle \cdots \rangle$ below) is proportional to the driving force parameter:
\begin{equation}
\label{equation:kappa}
\frac{\langle J^{\alpha} \rangle}{TV} = \sum_{\beta}\kappa^{\alpha\beta} F_{\rm e}^{\beta}.
\end{equation}
where $T$ is the system temperature, $V$ is the system volume, and $\kappa^{\alpha\beta}$ is the $\alpha\beta$-component of the \gls{ltc} tensor. For explicit expressions of the virial tensor and heat current vector applicable to the \gls{nep} model, we refer to the literature.~\cite{fan2015force,fan2021neuroevolution,Fan2022GPUMD} In this paper, the three \glspl{mof} have cubic symmetry and without loss of generality we can focus on the diagonal component in the $x$ direction, giving 
\begin{equation}
\label{equation:kappa_scaler}
\kappa^{xx} = \frac{\langle J^{x} \rangle}{TVF_{\rm e}^{x}}.
\end{equation}
The magnitude of the driving force parameter $F_{\rm e}$ needs to be small enough to keep the system within the linear-response regime but also large enough to induce a sufficiently large signal-to-noise ratio. We have tested that for all the three \glspl{mof} $F_{\rm e}^{x} =$ \SI{2e-4}{\per\angstrom} is an appropriate value for temperatures $\geq$ \SI{300}{K}, while a slightly smaller value of $F_{\rm e}^{x} =$ \SI{1.5e-4}{\per\angstrom} is more appropriate for lower temperatures. 

An appealing feature of the \gls{hnemd} method is that the \gls{ltc} can be spectrally decomposed \cite{Fan2019PRB}:
\begin{equation}
\kappa^{xx} = \int_0^{\infty} \frac{d\omega}{2\pi} \kappa^{xx}(\omega),
\end{equation}
where
\begin{equation}
\label{equation:kw}
\kappa^{xx}(\omega) = \frac{2}{VTF_{\rm e}^{x}} \int_{-\infty}^{\infty} dt e^{i\omega t} K^{x}(t).
\end{equation}
Here, $K^{x}(t)$ is the $x$-component of the virial-velocity correlation function:
\begin{equation}
\bm{K}(t) = \sum_i \langle \mathbf{W}_i(0) \cdot \bm{v}_i(t) \rangle.
\end{equation}
where $\bm{v}_i$ is the velocity of atom $i$.

\subsection{The \texorpdfstring{\gls{nemd}}{NEMD} method and spectral conductance in the ballistic regime}

With the \gls{hnemd} method alone, one can only access the diffusive transport properties. To obtain a more complete description of thermal transport from ballistic to diffusive, an established efficient scheme \cite{Fan2019PRB,li2019influence} is to supplement \gls{hnemd} with \gls{nemd} simulations. 

In our \gls{nemd} simulations, two local thermostats, i.e., heat source and heat sink, at different temperatures are used to generate a non-equilibrium steady state with constant heat flux. The thermal conductance $G$ between the two thermostats separated by distance $L$ can be calculated as:
\begin{equation}
\label{equation:G}
G^{xx}(L) = \frac{J^{x}}{\Delta T V},
\end{equation}
where $\Delta T$ is the temperature difference between the heat source and the heat sink. An effective \gls{ltc} (also called apparent \gls{ltc}) $\kappa^{xx} (L)$ for the finite system with a transport length of $L$ can be defined as \cite{li2019influence}:
\begin{equation}
\label{equation:GL}
\kappa^{xx} (L) = G^{xx}(L) L = \frac{J^{x}}{(\Delta T/L) V}.
\end{equation}
Similar to the \gls{ltc}, the thermal conductance can also be spectrally decomposed:
\begin{equation}
G^{xx}(L) = \int_0^{\infty} \frac{d\omega}{2\pi} G^{xx}(L,\omega),
\end{equation}
where
\begin{equation}
G^{xx}(L,\omega) = \frac{2}{V \Delta T} \int_{-\infty}^{\infty} dt e^{i\omega t} K^{x}(t).
\end{equation}
The limit of zero transport length $L\rightarrow 0$ corresponds to the ballistic limit $G^{xx}_{\rm b}(\omega)=G^{xx}(L\rightarrow 0,\omega)$, and one can define a spectrally decomposed \gls{mfp} as $\lambda^{xx}(\omega)=\kappa^{xx}(\omega)/G^{xx}_{\rm b}(\omega)$. With the spectral \gls{mfp}, we can obtain the spectral \gls{ltc} with any transport length
\begin{equation}
\label{equation:kwl}
\kappa^{xx} (L,\omega) = \frac{\kappa^{xx}(\omega)}{1 + \lambda^{xx}(\omega)/L},
\end{equation}
and the integrated \gls{ltc}
\begin{equation}
\kappa^{xx} (L) = \int_{0}^{\infty} \frac{d\omega}{2\pi} \kappa^{xx} (L, \omega) .
\end{equation}

\subsection{The \texorpdfstring{\gls{emd}}{EMD} method}

In the \gls{emd} method based on a Green-Kubo relation, \cite{green1954markoff, kubo1957statistical} the $xx$-component of the running \gls{ltc} can be expressed as an integral of the heat current auto-correlation function:
\begin{equation}
\label{equation:gk}
\kappa^{xx}(t) = \frac{1}{k_{\rm B} T^2V} \int_0^{t} dt' \langle J^{x}(0)J^{x}(t')\rangle,
\end{equation}
where $k_{\rm B}$ is Boltzmann's constant and $t$ is the upper limit of the correlation time.

\subsection{Phonon \texorpdfstring{\gls{sed}}{SED} calculations}

We perform the phonon \gls{sed} \cite{Thomas_PhysRevB_2010, Feng_2015} calculations to examine the phonon dispersion structures of \glspl{mof}. The phonon \gls{sed} is a function of wave vector ($\bm{k}$) and frequency ($\omega$), and can be calculated using the following formula in \gls{md} simulations:
\begin{align}
\Phi \left( \bm{k}, \omega \right) &= \frac{1}{4\pi\tau_{0}N}\sum_{\alpha}^{\{x,y,z\}} \sum_{b}^{B} m_{b} \nonumber \\
&\quad \times \Biggl\vert \int_{0}^{\tau_{0}}\sum_{n_{x,y,z}}^{N} v_{\alpha} 
\left(\begin{array}{c}
  n_{x,y,z} \\
  b
\end{array};t\right) \nonumber \\
&\quad \times \mathrm{exp} \left[ i\bm{k} \cdot \bm{r}
\left(\begin{array}{c}
  n_{x,y,z} \\
  0
\end{array}\right)- i\omega t\right] dt\Biggr\vert^{2},
\label{equation:sed}
\end{align}
where $\tau_{0}$ is the total simulation time, $N$ is the number of unit cells in the crystal, $b$ is the atom label in a given unit cell $n$, $m_{b}$ is the mass of atom $b$ in the unit cell, $n_{x,y,z}$ is the index number of unit cells along the $x$, $y$, and $z$ directions, $v_{\alpha} 
\left(\begin{array}{c}
n_{x,y,z} \\
  b
\end{array};t\right)$ denotes the velocity of atom $b$ of the $n$-th unit cell along the $\alpha$ direction at time $t$ and $\bm{r}
\left(\begin{array}{c}
  n_{x,y,z} \\
  0
\end{array}\right)$ is the equilibrium position of unit cell $n$.

\subsection{Details of the \texorpdfstring{\gls{md}}{MD} simulations}

We used the \textsc{gpumd} package (version 3.7) \cite{Fan2017CPC} to perform all the \gls{md} simulations. A time step of \SI{0.5}{\femto\second} was used with the velocity-Verlet integration scheme, which has been confirmed to be small enough. The time coupling parameter in the thermostat and barostat are \SI{50}{\femto\second} and \SI{500}{\femto\second}, respectively. In all \gls{hnemd}, \gls{emd}, and \gls{nemd} simulations, we calculated the statistical error of the \gls{ltc} as the standard error of the mean values from five independent runs. 

\subsubsection{\texorpdfstring{\gls{hnemd}}{HNEMD} simulations}

In \gls{hnemd} simulations, we first equilibrated for \SI{100}{\pico\second} in the isothermal-isobaric ($NpT$) ensemble at a given target temperature and zero pressure, followed by a \SI{2}{\nano\second} production stage in the $NVT$ ensemble. The $NpT$ ensemble was achieved by the Berendsen method \cite{berendsen1984molecular}. The $NVT$ ensemble was achieved by the Nose-Hoover chain thermostat  \cite{tuckerman2010statistical}. The external driving force was added during the production stage. 

\subsubsection{\texorpdfstring{\gls{nemd}}{NEMD} simulations}

In \gls{nemd} simulations, the system was only periodic in the lateral directions and the two ends in the transport direction were fixed. Next to one fixed end, a heat source region with a temperature of \SI{325}{K} was created by applying a Langevin thermostat \cite{bussi2007accurate} and a heat sink region with a temperature of \SI{275}{K} was similarly created next to the other fixed end. The thermostated regions contain two conventional cells in the transport direction which are sufficiently long to avoid artifacts \cite{li2019influence}. After an equilibration similar to the \gls{hnemd} simulation, the \gls{nemd} simulation with the two local Langevin thermostats was performed for \SI{1}{\nano\second}. All the simulated samples have \numproduct{5x5} conventional cells in the lateral directions and the lengths in the transport direction are varied to explore different transport regimes.

\subsubsection{\texorpdfstring{\gls{emd}}{EMD} simulations}
In \gls{emd} simulations, after an equilibration similar to the \gls{hnemd} simulation, a production run in the micro-canonical ($NVE$) ensemble was performed for \SI{10}{\nano\second}.

\subsubsection{Phonon \texorpdfstring{\gls{sed}}{SED} calculations}

For the phonon \gls{sed} calculations, we used a \numproduct{1x1x60} supercell for each \gls{mof}. All systems were first relaxed in the $NVT$ ensemble for \SI{0.5}{\nano\second}, followed by a \SI{0.75}{\nano\second} production stage in the $NVE$ ensemble. The velocities and positions of all atoms during the final \SI{0.5}{\nano\second} were collected to calculate the \gls{sed}.


\begin{thebibliography}{87}%
\makeatletter
\providecommand \@ifxundefined [1]{%
 \@ifx{#1\undefined}
}%
\providecommand \@ifnum [1]{%
 \ifnum #1\expandafter \@firstoftwo
 \else \expandafter \@secondoftwo
 \fi
}%
\providecommand \@ifx [1]{%
 \ifx #1\expandafter \@firstoftwo
 \else \expandafter \@secondoftwo
 \fi
}%
\providecommand \natexlab [1]{#1}%
\providecommand \enquote  [1]{``#1''}%
\providecommand \bibnamefont  [1]{#1}%
\providecommand \bibfnamefont [1]{#1}%
\providecommand \citenamefont [1]{#1}%
\providecommand \href@noop [0]{\@secondoftwo}%
\providecommand \href [0]{\begingroup \@sanitize@url \@href}%
\providecommand \@href[1]{\@@startlink{#1}\@@href}%
\providecommand \@@href[1]{\endgroup#1\@@endlink}%
\providecommand \@sanitize@url [0]{\catcode `\\12\catcode `\$12\catcode
  `\&12\catcode `\#12\catcode `\^12\catcode `\_12\catcode `\%12\relax}%
\providecommand \@@startlink[1]{}%
\providecommand \@@endlink[0]{}%
\providecommand \url  [0]{\begingroup\@sanitize@url \@url }%
\providecommand \@url [1]{\endgroup\@href {#1}{\urlprefix }}%
\providecommand \urlprefix  [0]{URL }%
\providecommand \Eprint [0]{\href }%
\providecommand \doibase [0]{https://doi.org/}%
\providecommand \selectlanguage [0]{\@gobble}%
\providecommand \bibinfo  [0]{\@secondoftwo}%
\providecommand \bibfield  [0]{\@secondoftwo}%
\providecommand \translation [1]{[#1]}%
\providecommand \BibitemOpen [0]{}%
\providecommand \bibitemStop [0]{}%
\providecommand \bibitemNoStop [0]{.\EOS\space}%
\providecommand \EOS [0]{\spacefactor3000\relax}%
\providecommand \BibitemShut  [1]{\csname bibitem#1\endcsname}%
\let\auto@bib@innerbib\@empty
\bibitem [{\citenamefont {Rowsell}\ and\ \citenamefont
  {Yaghi}(2004)}]{rowsell2004metal}%
  \BibitemOpen
  \bibfield  {author} {\bibinfo {author} {\bibfnamefont {J.~L.}\ \bibnamefont
  {Rowsell}}\ and\ \bibinfo {author} {\bibfnamefont {O.~M.}\ \bibnamefont
  {Yaghi}},\ }\bibfield  {title} {\bibinfo {title} {{Metal--organic frameworks:
  a new class of porous materials}},\ }\href
  {https://doi.org/10.1016/j.micromeso.2004.03.034} {\bibfield  {journal}
  {\bibinfo  {journal} {Microporous and Mesoporous Materials}\ }\textbf
  {\bibinfo {volume} {73}},\ \bibinfo {pages} {3} (\bibinfo {year}
  {2004})}\BibitemShut {NoStop}%
\bibitem [{\citenamefont {Yaghi}\ \emph {et~al.}(2003)\citenamefont {Yaghi},
  \citenamefont {O'Keeffe}, \citenamefont {Ockwig}, \citenamefont {Chae},
  \citenamefont {Eddaoudi},\ and\ \citenamefont {Kim}}]{yaghi2003reticular}%
  \BibitemOpen
  \bibfield  {author} {\bibinfo {author} {\bibfnamefont {O.~M.}\ \bibnamefont
  {Yaghi}}, \bibinfo {author} {\bibfnamefont {M.}~\bibnamefont {O'Keeffe}},
  \bibinfo {author} {\bibfnamefont {N.~W.}\ \bibnamefont {Ockwig}}, \bibinfo
  {author} {\bibfnamefont {H.~K.}\ \bibnamefont {Chae}}, \bibinfo {author}
  {\bibfnamefont {M.}~\bibnamefont {Eddaoudi}},\ and\ \bibinfo {author}
  {\bibfnamefont {J.}~\bibnamefont {Kim}},\ }\bibfield  {title} {\bibinfo
  {title} {{Reticular synthesis and the design of new materials}},\ }\href
  {https://doi.org/10.1038/nature01650} {\bibfield  {journal} {\bibinfo
  {journal} {Nature}\ }\textbf {\bibinfo {volume} {423}},\ \bibinfo {pages}
  {705} (\bibinfo {year} {2003})}\BibitemShut {NoStop}%
\bibitem [{\citenamefont {Li}\ \emph {et~al.}(2014)\citenamefont {Li},
  \citenamefont {Wen}, \citenamefont {Zhou},\ and\ \citenamefont
  {Chen}}]{li2014porous}%
  \BibitemOpen
  \bibfield  {author} {\bibinfo {author} {\bibfnamefont {B.}~\bibnamefont
  {Li}}, \bibinfo {author} {\bibfnamefont {H.-M.}\ \bibnamefont {Wen}},
  \bibinfo {author} {\bibfnamefont {W.}~\bibnamefont {Zhou}},\ and\ \bibinfo
  {author} {\bibfnamefont {B.}~\bibnamefont {Chen}},\ }\bibfield  {title}
  {\bibinfo {title} {{Porous metal--organic frameworks for gas storage and
  separation: what, how, and why?}},\ }\href
  {https://doi.org/10.1021/jz501586e} {\bibfield  {journal} {\bibinfo
  {journal} {The Journal of Physical Chemistry Letters}\ }\textbf {\bibinfo
  {volume} {5}},\ \bibinfo {pages} {3468} (\bibinfo {year} {2014})}\BibitemShut
  {NoStop}%
\bibitem [{\citenamefont {Hanikel}\ \emph {et~al.}(2020)\citenamefont
  {Hanikel}, \citenamefont {Pr{\'e}vot},\ and\ \citenamefont
  {Yaghi}}]{hanikel2020mof}%
  \BibitemOpen
  \bibfield  {author} {\bibinfo {author} {\bibfnamefont {N.}~\bibnamefont
  {Hanikel}}, \bibinfo {author} {\bibfnamefont {M.~S.}\ \bibnamefont
  {Pr{\'e}vot}},\ and\ \bibinfo {author} {\bibfnamefont {O.~M.}\ \bibnamefont
  {Yaghi}},\ }\bibfield  {title} {\bibinfo {title} {{MOF} water harvesters},\
  }\href {https://doi.org/10.1038/s41565-020-0673-x} {\bibfield  {journal}
  {\bibinfo  {journal} {Nature Nanotechnology}\ }\textbf {\bibinfo {volume}
  {15}},\ \bibinfo {pages} {348} (\bibinfo {year} {2020})}\BibitemShut
  {NoStop}%
\bibitem [{\citenamefont {Stassen}\ \emph {et~al.}(2017)\citenamefont
  {Stassen}, \citenamefont {Burtch}, \citenamefont {Talin}, \citenamefont
  {Falcaro}, \citenamefont {Allendorf},\ and\ \citenamefont
  {Ameloot}}]{stassen2017updated}%
  \BibitemOpen
  \bibfield  {author} {\bibinfo {author} {\bibfnamefont {I.}~\bibnamefont
  {Stassen}}, \bibinfo {author} {\bibfnamefont {N.}~\bibnamefont {Burtch}},
  \bibinfo {author} {\bibfnamefont {A.}~\bibnamefont {Talin}}, \bibinfo
  {author} {\bibfnamefont {P.}~\bibnamefont {Falcaro}}, \bibinfo {author}
  {\bibfnamefont {M.}~\bibnamefont {Allendorf}},\ and\ \bibinfo {author}
  {\bibfnamefont {R.}~\bibnamefont {Ameloot}},\ }\bibfield  {title} {\bibinfo
  {title} {{An updated roadmap for the integration of metal--organic frameworks
  with electronic devices and chemical sensors}},\ }\href
  {https://doi.org/10.1039/C7CS00122C} {\bibfield  {journal} {\bibinfo
  {journal} {Chemical Society Reviews}\ }\textbf {\bibinfo {volume} {46}},\
  \bibinfo {pages} {3185} (\bibinfo {year} {2017})}\BibitemShut {NoStop}%
\bibitem [{\citenamefont {Rogge}\ \emph {et~al.}(2017)\citenamefont {Rogge},
  \citenamefont {Bavykina}, \citenamefont {Hajek}, \citenamefont {Garcia},
  \citenamefont {Olivos-Suarez}, \citenamefont {Sep{\'u}lveda-Escribano},
  \citenamefont {Vimont}, \citenamefont {Clet}, \citenamefont {Bazin},
  \citenamefont {Kapteijn} \emph {et~al.}}]{rogge2017metal}%
  \BibitemOpen
  \bibfield  {author} {\bibinfo {author} {\bibfnamefont {S.~M.}\ \bibnamefont
  {Rogge}}, \bibinfo {author} {\bibfnamefont {A.}~\bibnamefont {Bavykina}},
  \bibinfo {author} {\bibfnamefont {J.}~\bibnamefont {Hajek}}, \bibinfo
  {author} {\bibfnamefont {H.}~\bibnamefont {Garcia}}, \bibinfo {author}
  {\bibfnamefont {A.~I.}\ \bibnamefont {Olivos-Suarez}}, \bibinfo {author}
  {\bibfnamefont {A.}~\bibnamefont {Sep{\'u}lveda-Escribano}}, \bibinfo
  {author} {\bibfnamefont {A.}~\bibnamefont {Vimont}}, \bibinfo {author}
  {\bibfnamefont {G.}~\bibnamefont {Clet}}, \bibinfo {author} {\bibfnamefont
  {P.}~\bibnamefont {Bazin}}, \bibinfo {author} {\bibfnamefont
  {F.}~\bibnamefont {Kapteijn}}, \emph {et~al.},\ }\bibfield  {title} {\bibinfo
  {title} {{Metal--organic and covalent organic frameworks as single-site
  catalysts}},\ }\href {https://doi.org/10.1039/C7CS00033B} {\bibfield
  {journal} {\bibinfo  {journal} {Chemical Society Reviews}\ }\textbf {\bibinfo
  {volume} {46}},\ \bibinfo {pages} {3134} (\bibinfo {year}
  {2017})}\BibitemShut {NoStop}%
\bibitem [{\citenamefont {Huang}\ \emph
  {et~al.}(2007{\natexlab{a}})\citenamefont {Huang}, \citenamefont {Ni},
  \citenamefont {Millward}, \citenamefont {McGaughey}, \citenamefont {Uher},
  \citenamefont {Kaviany},\ and\ \citenamefont {Yaghi}}]{huang2007thermal2}%
  \BibitemOpen
  \bibfield  {author} {\bibinfo {author} {\bibfnamefont {B.}~\bibnamefont
  {Huang}}, \bibinfo {author} {\bibfnamefont {Z.}~\bibnamefont {Ni}}, \bibinfo
  {author} {\bibfnamefont {A.}~\bibnamefont {Millward}}, \bibinfo {author}
  {\bibfnamefont {A.}~\bibnamefont {McGaughey}}, \bibinfo {author}
  {\bibfnamefont {C.}~\bibnamefont {Uher}}, \bibinfo {author} {\bibfnamefont
  {M.}~\bibnamefont {Kaviany}},\ and\ \bibinfo {author} {\bibfnamefont
  {O.}~\bibnamefont {Yaghi}},\ }\bibfield  {title} {\bibinfo {title} {{Thermal
  conductivity of a metal-organic framework (MOF-5): Part II. Measurement}},\
  }\href {https://doi.org/10.1016/j.ijheatmasstransfer.2006.10.001} {\bibfield
  {journal} {\bibinfo  {journal} {International Journal of Heat and Mass
  Transfer}\ }\textbf {\bibinfo {volume} {50}},\ \bibinfo {pages} {405}
  (\bibinfo {year} {2007}{\natexlab{a}})}\BibitemShut {NoStop}%
\bibitem [{\citenamefont {Ming}\ \emph {et~al.}(2014)\citenamefont {Ming},
  \citenamefont {Purewal}, \citenamefont {Sudik}, \citenamefont {Xu},
  \citenamefont {Yang}, \citenamefont {Veenstra}, \citenamefont {Rhodes},
  \citenamefont {Soltis}, \citenamefont {Warner}, \citenamefont {Gaab} \emph
  {et~al.}}]{ming2014thermophysical}%
  \BibitemOpen
  \bibfield  {author} {\bibinfo {author} {\bibfnamefont {Y.}~\bibnamefont
  {Ming}}, \bibinfo {author} {\bibfnamefont {J.}~\bibnamefont {Purewal}},
  \bibinfo {author} {\bibfnamefont {A.}~\bibnamefont {Sudik}}, \bibinfo
  {author} {\bibfnamefont {C.}~\bibnamefont {Xu}}, \bibinfo {author}
  {\bibfnamefont {J.}~\bibnamefont {Yang}}, \bibinfo {author} {\bibfnamefont
  {M.}~\bibnamefont {Veenstra}}, \bibinfo {author} {\bibfnamefont
  {K.}~\bibnamefont {Rhodes}}, \bibinfo {author} {\bibfnamefont
  {R.}~\bibnamefont {Soltis}}, \bibinfo {author} {\bibfnamefont
  {J.}~\bibnamefont {Warner}}, \bibinfo {author} {\bibfnamefont
  {M.}~\bibnamefont {Gaab}}, \emph {et~al.},\ }\bibfield  {title} {\bibinfo
  {title} {Thermophysical properties of {MOF}-5 powders},\ }\href
  {https://doi.org/10.1016/j.micromeso.2013.11.015} {\bibfield  {journal}
  {\bibinfo  {journal} {Microporous and Mesoporous Materials}\ }\textbf
  {\bibinfo {volume} {185}},\ \bibinfo {pages} {235} (\bibinfo {year}
  {2014})}\BibitemShut {NoStop}%
\bibitem [{\citenamefont {Erickson}\ \emph {et~al.}(2015)\citenamefont
  {Erickson}, \citenamefont {L{\'e}onard}, \citenamefont {Stavila},
  \citenamefont {Foster}, \citenamefont {Spataru}, \citenamefont {Jones},
  \citenamefont {Foley}, \citenamefont {Hopkins}, \citenamefont {Allendorf},\
  and\ \citenamefont {Talin}}]{erickson2015thin}%
  \BibitemOpen
  \bibfield  {author} {\bibinfo {author} {\bibfnamefont {K.~J.}\ \bibnamefont
  {Erickson}}, \bibinfo {author} {\bibfnamefont {F.}~\bibnamefont
  {L{\'e}onard}}, \bibinfo {author} {\bibfnamefont {V.}~\bibnamefont
  {Stavila}}, \bibinfo {author} {\bibfnamefont {M.~E.}\ \bibnamefont {Foster}},
  \bibinfo {author} {\bibfnamefont {C.~D.}\ \bibnamefont {Spataru}}, \bibinfo
  {author} {\bibfnamefont {R.~E.}\ \bibnamefont {Jones}}, \bibinfo {author}
  {\bibfnamefont {B.~M.}\ \bibnamefont {Foley}}, \bibinfo {author}
  {\bibfnamefont {P.~E.}\ \bibnamefont {Hopkins}}, \bibinfo {author}
  {\bibfnamefont {M.~D.}\ \bibnamefont {Allendorf}},\ and\ \bibinfo {author}
  {\bibfnamefont {A.~A.}\ \bibnamefont {Talin}},\ }\bibfield  {title} {\bibinfo
  {title} {Thin film thermoelectric metal--organic framework with high seebeck
  coefficient and low thermal conductivity},\ }\href
  {https://doi.org/10.1002/adma.201501078} {\bibfield  {journal} {\bibinfo
  {journal} {Advanced Materials}\ }\textbf {\bibinfo {volume} {27}},\ \bibinfo
  {pages} {3453} (\bibinfo {year} {2015})}\BibitemShut {NoStop}%
\bibitem [{\citenamefont {Semelsberger}\ \emph {et~al.}(2016)\citenamefont
  {Semelsberger}, \citenamefont {Veenstra},\ and\ \citenamefont
  {Dixon}}]{semelsberger2016room}%
  \BibitemOpen
  \bibfield  {author} {\bibinfo {author} {\bibfnamefont {T.~A.}\ \bibnamefont
  {Semelsberger}}, \bibinfo {author} {\bibfnamefont {M.}~\bibnamefont
  {Veenstra}},\ and\ \bibinfo {author} {\bibfnamefont {C.}~\bibnamefont
  {Dixon}},\ }\bibfield  {title} {\bibinfo {title} {{Room temperature thermal
  conductivity measurements of neat MOF-5 compacts with high pressure hydrogen
  and helium}},\ }\href {https://doi.org/10.1016/j.ijhydene.2015.12.059}
  {\bibfield  {journal} {\bibinfo  {journal} {International Journal of Hydrogen
  Energy}\ }\textbf {\bibinfo {volume} {41}},\ \bibinfo {pages} {4690}
  (\bibinfo {year} {2016})}\BibitemShut {NoStop}%
\bibitem [{\citenamefont {Gunatilleke}\ \emph {et~al.}(2017)\citenamefont
  {Gunatilleke}, \citenamefont {Wei}, \citenamefont {Niu}, \citenamefont
  {Wojtas}, \citenamefont {Nolas},\ and\ \citenamefont
  {Ma}}]{gunatilleke2017thermal}%
  \BibitemOpen
  \bibfield  {author} {\bibinfo {author} {\bibfnamefont {W.~D.}\ \bibnamefont
  {Gunatilleke}}, \bibinfo {author} {\bibfnamefont {K.}~\bibnamefont {Wei}},
  \bibinfo {author} {\bibfnamefont {Z.}~\bibnamefont {Niu}}, \bibinfo {author}
  {\bibfnamefont {L.}~\bibnamefont {Wojtas}}, \bibinfo {author} {\bibfnamefont
  {G.}~\bibnamefont {Nolas}},\ and\ \bibinfo {author} {\bibfnamefont
  {S.}~\bibnamefont {Ma}},\ }\bibfield  {title} {\bibinfo {title} {Thermal
  conductivity of a perovskite-type metal--organic framework crystal},\ }\href
  {https://doi.org/10.1039/C7DT02927F} {\bibfield  {journal} {\bibinfo
  {journal} {Dalton Transactions}\ }\textbf {\bibinfo {volume} {46}},\ \bibinfo
  {pages} {13342} (\bibinfo {year} {2017})}\BibitemShut {NoStop}%
\bibitem [{\citenamefont {Sun}\ \emph {et~al.}(2017)\citenamefont {Sun},
  \citenamefont {Liao}, \citenamefont {Sheberla}, \citenamefont {Kraemer},
  \citenamefont {Zhou}, \citenamefont {Stach}, \citenamefont {Zakharov},
  \citenamefont {Stavila}, \citenamefont {Talin}, \citenamefont {Ge} \emph
  {et~al.}}]{sun2017microporous}%
  \BibitemOpen
  \bibfield  {author} {\bibinfo {author} {\bibfnamefont {L.}~\bibnamefont
  {Sun}}, \bibinfo {author} {\bibfnamefont {B.}~\bibnamefont {Liao}}, \bibinfo
  {author} {\bibfnamefont {D.}~\bibnamefont {Sheberla}}, \bibinfo {author}
  {\bibfnamefont {D.}~\bibnamefont {Kraemer}}, \bibinfo {author} {\bibfnamefont
  {J.}~\bibnamefont {Zhou}}, \bibinfo {author} {\bibfnamefont {E.~A.}\
  \bibnamefont {Stach}}, \bibinfo {author} {\bibfnamefont {D.}~\bibnamefont
  {Zakharov}}, \bibinfo {author} {\bibfnamefont {V.}~\bibnamefont {Stavila}},
  \bibinfo {author} {\bibfnamefont {A.~A.}\ \bibnamefont {Talin}}, \bibinfo
  {author} {\bibfnamefont {Y.}~\bibnamefont {Ge}}, \emph {et~al.},\ }\bibfield
  {title} {\bibinfo {title} {A microporous and naturally nanostructured
  thermoelectric metal-organic framework with ultralow thermal conductivity},\
  }\href {https://doi.org/10.1016/j.joule.2017.07.018} {\bibfield  {journal}
  {\bibinfo  {journal} {Joule}\ }\textbf {\bibinfo {volume} {1}},\ \bibinfo
  {pages} {168} (\bibinfo {year} {2017})}\BibitemShut {NoStop}%
\bibitem [{\citenamefont {Huang}\ \emph {et~al.}(2019)\citenamefont {Huang},
  \citenamefont {Xia}, \citenamefont {Hu}, \citenamefont {Li},\ and\
  \citenamefont {Liu}}]{huang2019general}%
  \BibitemOpen
  \bibfield  {author} {\bibinfo {author} {\bibfnamefont {J.}~\bibnamefont
  {Huang}}, \bibinfo {author} {\bibfnamefont {X.}~\bibnamefont {Xia}}, \bibinfo
  {author} {\bibfnamefont {X.}~\bibnamefont {Hu}}, \bibinfo {author}
  {\bibfnamefont {S.}~\bibnamefont {Li}},\ and\ \bibinfo {author}
  {\bibfnamefont {K.}~\bibnamefont {Liu}},\ }\bibfield  {title} {\bibinfo
  {title} {A general method for measuring the thermal conductivity of mof
  crystals},\ }\href {https://doi.org/10.1016/j.ijheatmasstransfer.2019.04.018}
  {\bibfield  {journal} {\bibinfo  {journal} {International Journal of Heat and
  Mass Transfer}\ }\textbf {\bibinfo {volume} {138}},\ \bibinfo {pages} {11}
  (\bibinfo {year} {2019})}\BibitemShut {NoStop}%
\bibitem [{\citenamefont {Huang}\ \emph {et~al.}(2020)\citenamefont {Huang},
  \citenamefont {Fan}, \citenamefont {Xia}, \citenamefont {Li},\ and\
  \citenamefont {Zhang}}]{huang2020situ}%
  \BibitemOpen
  \bibfield  {author} {\bibinfo {author} {\bibfnamefont {J.}~\bibnamefont
  {Huang}}, \bibinfo {author} {\bibfnamefont {A.}~\bibnamefont {Fan}}, \bibinfo
  {author} {\bibfnamefont {X.}~\bibnamefont {Xia}}, \bibinfo {author}
  {\bibfnamefont {S.}~\bibnamefont {Li}},\ and\ \bibinfo {author}
  {\bibfnamefont {X.}~\bibnamefont {Zhang}},\ }\bibfield  {title} {\bibinfo
  {title} {{In situ thermal vonductivity measurement of single-crystal zeolitic
  imidazolate framework-8 by raman-resistance temperature detectors method}},\
  }\href {https://doi.org/10.1021/acsnano.0c06756} {\bibfield  {journal}
  {\bibinfo  {journal} {ACS Nano}\ }\textbf {\bibinfo {volume} {14}},\ \bibinfo
  {pages} {14100} (\bibinfo {year} {2020})}\BibitemShut {NoStop}%
\bibitem [{\citenamefont {Babaei}\ \emph
  {et~al.}(2020{\natexlab{a}})\citenamefont {Babaei}, \citenamefont {DeCoster},
  \citenamefont {Jeong}, \citenamefont {Hassan}, \citenamefont {Islamoglu},
  \citenamefont {Baumgart}, \citenamefont {McGaughey}, \citenamefont {Redel},
  \citenamefont {Farha}, \citenamefont {Hopkins} \emph
  {et~al.}}]{babaei2020observation}%
  \BibitemOpen
  \bibfield  {author} {\bibinfo {author} {\bibfnamefont {H.}~\bibnamefont
  {Babaei}}, \bibinfo {author} {\bibfnamefont {M.~E.}\ \bibnamefont
  {DeCoster}}, \bibinfo {author} {\bibfnamefont {M.}~\bibnamefont {Jeong}},
  \bibinfo {author} {\bibfnamefont {Z.~M.}\ \bibnamefont {Hassan}}, \bibinfo
  {author} {\bibfnamefont {T.}~\bibnamefont {Islamoglu}}, \bibinfo {author}
  {\bibfnamefont {H.}~\bibnamefont {Baumgart}}, \bibinfo {author}
  {\bibfnamefont {A.~J.}\ \bibnamefont {McGaughey}}, \bibinfo {author}
  {\bibfnamefont {E.}~\bibnamefont {Redel}}, \bibinfo {author} {\bibfnamefont
  {O.~K.}\ \bibnamefont {Farha}}, \bibinfo {author} {\bibfnamefont {P.~E.}\
  \bibnamefont {Hopkins}}, \emph {et~al.},\ }\bibfield  {title} {\bibinfo
  {title} {Observation of reduced thermal conductivity in a metal-organic
  framework due to the presence of adsorbates},\ }\href
  {https://doi.org/10.1038/s41467-020-17822-0} {\bibfield  {journal} {\bibinfo
  {journal} {Nature Communications}\ }\textbf {\bibinfo {volume} {11}},\
  \bibinfo {pages} {1} (\bibinfo {year} {2020}{\natexlab{a}})}\BibitemShut
  {NoStop}%
\bibitem [{\citenamefont {Huang}\ \emph
  {et~al.}(2007{\natexlab{b}})\citenamefont {Huang}, \citenamefont
  {McGaughey},\ and\ \citenamefont {Kaviany}}]{huang2007thermal1}%
  \BibitemOpen
  \bibfield  {author} {\bibinfo {author} {\bibfnamefont {B.}~\bibnamefont
  {Huang}}, \bibinfo {author} {\bibfnamefont {A.}~\bibnamefont {McGaughey}},\
  and\ \bibinfo {author} {\bibfnamefont {M.}~\bibnamefont {Kaviany}},\
  }\bibfield  {title} {\bibinfo {title} {{Thermal conductivity of metal-organic
  framework 5 (MOF-5): Part I. Molecular dynamics simulations}},\ }\href
  {https://doi.org/10.1016/j.ijheatmasstransfer.2006.10.002} {\bibfield
  {journal} {\bibinfo  {journal} {International Journal of Heat and Mass
  Transfer}\ }\textbf {\bibinfo {volume} {50}},\ \bibinfo {pages} {393}
  (\bibinfo {year} {2007}{\natexlab{b}})}\BibitemShut {NoStop}%
\bibitem [{\citenamefont {Zhang}\ and\ \citenamefont
  {Jiang}(2013)}]{zhang2013thermal}%
  \BibitemOpen
  \bibfield  {author} {\bibinfo {author} {\bibfnamefont {X.}~\bibnamefont
  {Zhang}}\ and\ \bibinfo {author} {\bibfnamefont {J.}~\bibnamefont {Jiang}},\
  }\bibfield  {title} {\bibinfo {title} {Thermal conductivity of zeolitic
  imidazolate framework-8: A molecular simulation study},\ }\href
  {https://doi.org/10.1021/jp405156y} {\bibfield  {journal} {\bibinfo
  {journal} {The Journal of Physical Chemistry C}\ }\textbf {\bibinfo {volume}
  {117}},\ \bibinfo {pages} {18441} (\bibinfo {year} {2013})}\BibitemShut
  {NoStop}%
\bibitem [{\citenamefont {Han}\ \emph {et~al.}(2014)\citenamefont {Han},
  \citenamefont {Budge},\ and\ \citenamefont {Greaney}}]{han2014relationship}%
  \BibitemOpen
  \bibfield  {author} {\bibinfo {author} {\bibfnamefont {L.}~\bibnamefont
  {Han}}, \bibinfo {author} {\bibfnamefont {M.}~\bibnamefont {Budge}},\ and\
  \bibinfo {author} {\bibfnamefont {P.~A.}\ \bibnamefont {Greaney}},\
  }\bibfield  {title} {\bibinfo {title} {Relationship between thermal
  conductivity and framework architecture in {MOF}-5},\ }\href
  {https://doi.org/10.1016/j.commatsci.2014.06.008} {\bibfield  {journal}
  {\bibinfo  {journal} {Computational Materials Science}\ }\textbf {\bibinfo
  {volume} {94}},\ \bibinfo {pages} {292} (\bibinfo {year} {2014})}\BibitemShut
  {NoStop}%
\bibitem [{\citenamefont {Babaei}\ and\ \citenamefont
  {Wilmer}(2016)}]{babaei2016mechanisms}%
  \BibitemOpen
  \bibfield  {author} {\bibinfo {author} {\bibfnamefont {H.}~\bibnamefont
  {Babaei}}\ and\ \bibinfo {author} {\bibfnamefont {C.~E.}\ \bibnamefont
  {Wilmer}},\ }\bibfield  {title} {\bibinfo {title} {Mechanisms of heat
  transfer in porous crystals containing adsorbed gases: Applications to
  metal-organic frameworks},\ }\href
  {https://doi.org/10.1103/PhysRevLett.116.025902} {\bibfield  {journal}
  {\bibinfo  {journal} {Physical Review Letters}\ }\textbf {\bibinfo {volume}
  {116}},\ \bibinfo {pages} {025902} (\bibinfo {year} {2016})}\BibitemShut
  {NoStop}%
\bibitem [{\citenamefont {Babaei}\ \emph {et~al.}(2017)\citenamefont {Babaei},
  \citenamefont {McGaughey},\ and\ \citenamefont {Wilmer}}]{babaei2017effect}%
  \BibitemOpen
  \bibfield  {author} {\bibinfo {author} {\bibfnamefont {H.}~\bibnamefont
  {Babaei}}, \bibinfo {author} {\bibfnamefont {A.~J.}\ \bibnamefont
  {McGaughey}},\ and\ \bibinfo {author} {\bibfnamefont {C.~E.}\ \bibnamefont
  {Wilmer}},\ }\bibfield  {title} {\bibinfo {title} {Effect of pore size and
  shape on the thermal conductivity of metal-organic frameworks},\ }\href
  {https://doi.org/10.1039/C6SC03704F} {\bibfield  {journal} {\bibinfo
  {journal} {Chemical Science}\ }\textbf {\bibinfo {volume} {8}},\ \bibinfo
  {pages} {583} (\bibinfo {year} {2017})}\BibitemShut {NoStop}%
\bibitem [{\citenamefont {Sezginel}\ \emph {et~al.}(2018)\citenamefont
  {Sezginel}, \citenamefont {Asinger}, \citenamefont {Babaei},\ and\
  \citenamefont {Wilmer}}]{sezginel2018thermal}%
  \BibitemOpen
  \bibfield  {author} {\bibinfo {author} {\bibfnamefont {K.~B.}\ \bibnamefont
  {Sezginel}}, \bibinfo {author} {\bibfnamefont {P.~A.}\ \bibnamefont
  {Asinger}}, \bibinfo {author} {\bibfnamefont {H.}~\bibnamefont {Babaei}},\
  and\ \bibinfo {author} {\bibfnamefont {C.~E.}\ \bibnamefont {Wilmer}},\
  }\bibfield  {title} {\bibinfo {title} {Thermal transport in interpenetrated
  metal--organic frameworks},\ }\href
  {https://doi.org/10.1021/acs.chemmater.7b05015} {\bibfield  {journal}
  {\bibinfo  {journal} {Chemistry of Materials}\ }\textbf {\bibinfo {volume}
  {30}},\ \bibinfo {pages} {2281} (\bibinfo {year} {2018})}\BibitemShut
  {NoStop}%
\bibitem [{\citenamefont {Wieme}\ \emph {et~al.}(2019)\citenamefont {Wieme},
  \citenamefont {Vandenbrande}, \citenamefont {Lamaire}, \citenamefont {Kapil},
  \citenamefont {Vanduyfhuys},\ and\ \citenamefont
  {Van~Speybroeck}}]{wieme2019thermal}%
  \BibitemOpen
  \bibfield  {author} {\bibinfo {author} {\bibfnamefont {J.}~\bibnamefont
  {Wieme}}, \bibinfo {author} {\bibfnamefont {S.}~\bibnamefont {Vandenbrande}},
  \bibinfo {author} {\bibfnamefont {A.}~\bibnamefont {Lamaire}}, \bibinfo
  {author} {\bibfnamefont {V.}~\bibnamefont {Kapil}}, \bibinfo {author}
  {\bibfnamefont {L.}~\bibnamefont {Vanduyfhuys}},\ and\ \bibinfo {author}
  {\bibfnamefont {V.}~\bibnamefont {Van~Speybroeck}},\ }\bibfield  {title}
  {\bibinfo {title} {Thermal engineering of metal--organic frameworks for
  adsorption applications: a molecular simulation perspective},\ }\href
  {https://doi.org/10.1021/acsami.9b12533} {\bibfield  {journal} {\bibinfo
  {journal} {ACS Applied Materials \& Interfaces}\ }\textbf {\bibinfo {volume}
  {11}},\ \bibinfo {pages} {38697} (\bibinfo {year} {2019})}\BibitemShut
  {NoStop}%
\bibitem [{\citenamefont {Wei}\ \emph {et~al.}(2019)\citenamefont {Wei},
  \citenamefont {Huang}, \citenamefont {Li}, \citenamefont {Peng},\ and\
  \citenamefont {Li}}]{wei2019impacts}%
  \BibitemOpen
  \bibfield  {author} {\bibinfo {author} {\bibfnamefont {W.}~\bibnamefont
  {Wei}}, \bibinfo {author} {\bibfnamefont {J.}~\bibnamefont {Huang}}, \bibinfo
  {author} {\bibfnamefont {W.}~\bibnamefont {Li}}, \bibinfo {author}
  {\bibfnamefont {H.}~\bibnamefont {Peng}},\ and\ \bibinfo {author}
  {\bibfnamefont {S.}~\bibnamefont {Li}},\ }\bibfield  {title} {\bibinfo
  {title} {Impacts of ethanol and water adsorptions on thermal conductivity of
  {ZIF-8}},\ }\href {https://doi.org/10.1021/acs.jpcc.9b08187} {\bibfield
  {journal} {\bibinfo  {journal} {The Journal of Physical Chemistry C}\
  }\textbf {\bibinfo {volume} {123}},\ \bibinfo {pages} {27369} (\bibinfo
  {year} {2019})}\BibitemShut {NoStop}%
\bibitem [{\citenamefont {S{\o}rensen}\ \emph {et~al.}(2020)\citenamefont
  {S{\o}rensen}, \citenamefont {{\O}stergaard}, \citenamefont {Stepniewska},
  \citenamefont {Johra}, \citenamefont {Yue},\ and\ \citenamefont
  {Smedskjaer}}]{sorensen2020metal}%
  \BibitemOpen
  \bibfield  {author} {\bibinfo {author} {\bibfnamefont {S.~S.}\ \bibnamefont
  {S{\o}rensen}}, \bibinfo {author} {\bibfnamefont {M.~B.}\ \bibnamefont
  {{\O}stergaard}}, \bibinfo {author} {\bibfnamefont {M.}~\bibnamefont
  {Stepniewska}}, \bibinfo {author} {\bibfnamefont {H.}~\bibnamefont {Johra}},
  \bibinfo {author} {\bibfnamefont {Y.}~\bibnamefont {Yue}},\ and\ \bibinfo
  {author} {\bibfnamefont {M.~M.}\ \bibnamefont {Smedskjaer}},\ }\bibfield
  {title} {\bibinfo {title} {{Metal--organic framework glasses possess higher
  thermal conductivity than their crystalline counterparts}},\ }\href
  {https://doi.org/10.1021/acsami.0c02310} {\bibfield  {journal} {\bibinfo
  {journal} {ACS Applied Materials \& Interfaces}\ }\textbf {\bibinfo {volume}
  {12}},\ \bibinfo {pages} {18893} (\bibinfo {year} {2020})}\BibitemShut
  {NoStop}%
\bibitem [{\citenamefont {Islamov}\ \emph {et~al.}(2020)\citenamefont
  {Islamov}, \citenamefont {Babaei},\ and\ \citenamefont
  {Wilmer}}]{islamov2020influence}%
  \BibitemOpen
  \bibfield  {author} {\bibinfo {author} {\bibfnamefont {M.}~\bibnamefont
  {Islamov}}, \bibinfo {author} {\bibfnamefont {H.}~\bibnamefont {Babaei}},\
  and\ \bibinfo {author} {\bibfnamefont {C.~E.}\ \bibnamefont {Wilmer}},\
  }\bibfield  {title} {\bibinfo {title} {Influence of missing linker defects on
  the thermal conductivity of metal--organic framework {HKUST-1}},\ }\href
  {https://doi.org/10.1021/acsami.0c16127} {\bibfield  {journal} {\bibinfo
  {journal} {ACS Applied Materials \& Interfaces}\ }\textbf {\bibinfo {volume}
  {12}},\ \bibinfo {pages} {56172} (\bibinfo {year} {2020})}\BibitemShut
  {NoStop}%
\bibitem [{\citenamefont {Sezginel}\ \emph {et~al.}(2020)\citenamefont
  {Sezginel}, \citenamefont {Lee}, \citenamefont {Babaei},\ and\ \citenamefont
  {Wilmer}}]{sezginel2020effect}%
  \BibitemOpen
  \bibfield  {author} {\bibinfo {author} {\bibfnamefont {K.~B.}\ \bibnamefont
  {Sezginel}}, \bibinfo {author} {\bibfnamefont {S.}~\bibnamefont {Lee}},
  \bibinfo {author} {\bibfnamefont {H.}~\bibnamefont {Babaei}},\ and\ \bibinfo
  {author} {\bibfnamefont {C.~E.}\ \bibnamefont {Wilmer}},\ }\bibfield  {title}
  {\bibinfo {title} {Effect of flexibility on thermal transport in breathing
  porous crystals},\ }\href {https://doi.org/10.1021/acs.jpcc.0c04353}
  {\bibfield  {journal} {\bibinfo  {journal} {The Journal of Physical Chemistry
  C}\ }\textbf {\bibinfo {volume} {124}},\ \bibinfo {pages} {18604} (\bibinfo
  {year} {2020})}\BibitemShut {NoStop}%
\bibitem [{\citenamefont {Babaei}\ \emph
  {et~al.}(2020{\natexlab{b}})\citenamefont {Babaei}, \citenamefont {Lee},
  \citenamefont {Dods}, \citenamefont {Wilmer},\ and\ \citenamefont
  {Long}}]{babaei2020enhanced}%
  \BibitemOpen
  \bibfield  {author} {\bibinfo {author} {\bibfnamefont {H.}~\bibnamefont
  {Babaei}}, \bibinfo {author} {\bibfnamefont {J.-H.}\ \bibnamefont {Lee}},
  \bibinfo {author} {\bibfnamefont {M.~N.}\ \bibnamefont {Dods}}, \bibinfo
  {author} {\bibfnamefont {C.~E.}\ \bibnamefont {Wilmer}},\ and\ \bibinfo
  {author} {\bibfnamefont {J.~R.}\ \bibnamefont {Long}},\ }\bibfield  {title}
  {\bibinfo {title} {Enhanced thermal conductivity in a diamine-appended
  metal--organic framework as a result of cooperative {CO}$_2$ adsorption},\
  }\href {https://doi.org/10.1021/acsami.0c10233} {\bibfield  {journal}
  {\bibinfo  {journal} {ACS Applied materials \& Interfaces}\ }\textbf
  {\bibinfo {volume} {12}},\ \bibinfo {pages} {44617} (\bibinfo {year}
  {2020}{\natexlab{b}})}\BibitemShut {NoStop}%
\bibitem [{\citenamefont {Ying}\ \emph {et~al.}(2020)\citenamefont {Ying},
  \citenamefont {Zhang}, \citenamefont {Zhang},\ and\ \citenamefont
  {Zhong}}]{ying2020impacts}%
  \BibitemOpen
  \bibfield  {author} {\bibinfo {author} {\bibfnamefont {P.}~\bibnamefont
  {Ying}}, \bibinfo {author} {\bibfnamefont {J.}~\bibnamefont {Zhang}},
  \bibinfo {author} {\bibfnamefont {X.}~\bibnamefont {Zhang}},\ and\ \bibinfo
  {author} {\bibfnamefont {Z.}~\bibnamefont {Zhong}},\ }\bibfield  {title}
  {\bibinfo {title} {Impacts of functional group substitution and pressure on
  the thermal conductivity of {ZIF-8}},\ }\href
  {https://doi.org/10.1021/acs.jpcc.0c00597} {\bibfield  {journal} {\bibinfo
  {journal} {The Journal of Physical Chemistry C}\ }\textbf {\bibinfo {volume}
  {124}},\ \bibinfo {pages} {6274} (\bibinfo {year} {2020})}\BibitemShut
  {NoStop}%
\bibitem [{\citenamefont {Ying}\ \emph {et~al.}(2021)\citenamefont {Ying},
  \citenamefont {Zhang},\ and\ \citenamefont {Zhong}}]{ying2021effect}%
  \BibitemOpen
  \bibfield  {author} {\bibinfo {author} {\bibfnamefont {P.}~\bibnamefont
  {Ying}}, \bibinfo {author} {\bibfnamefont {J.}~\bibnamefont {Zhang}},\ and\
  \bibinfo {author} {\bibfnamefont {Z.}~\bibnamefont {Zhong}},\ }\bibfield
  {title} {\bibinfo {title} {Effect of phase transition on the thermal
  transport in isoreticular dut materials},\ }\href
  {https://doi.org/10.1021/acs.jpcc.1c02767} {\bibfield  {journal} {\bibinfo
  {journal} {The Journal of Physical Chemistry C}\ }\textbf {\bibinfo {volume}
  {125}},\ \bibinfo {pages} {12991} (\bibinfo {year} {2021})}\BibitemShut
  {NoStop}%
\bibitem [{\citenamefont {Cheng}\ \emph {et~al.}(2021)\citenamefont {Cheng},
  \citenamefont {Li}, \citenamefont {Wei}, \citenamefont {Huang},\ and\
  \citenamefont {Li}}]{cheng2021molecular}%
  \BibitemOpen
  \bibfield  {author} {\bibinfo {author} {\bibfnamefont {R.}~\bibnamefont
  {Cheng}}, \bibinfo {author} {\bibfnamefont {W.}~\bibnamefont {Li}}, \bibinfo
  {author} {\bibfnamefont {W.}~\bibnamefont {Wei}}, \bibinfo {author}
  {\bibfnamefont {J.}~\bibnamefont {Huang}},\ and\ \bibinfo {author}
  {\bibfnamefont {S.}~\bibnamefont {Li}},\ }\bibfield  {title} {\bibinfo
  {title} {Molecular insights into the correlation between microstructure and
  thermal conductivity of zeolitic imidazolate frameworks},\ }\href
  {https://doi.org/10.1021/acsami.0c21220} {\bibfield  {journal} {\bibinfo
  {journal} {ACS Applied Materials \& Interfaces}\ }\textbf {\bibinfo {volume}
  {13}},\ \bibinfo {pages} {14141} (\bibinfo {year} {2021})}\BibitemShut
  {NoStop}%
\bibitem [{\citenamefont {Zhang}\ \emph {et~al.}(2021)\citenamefont {Zhang},
  \citenamefont {Liu},\ and\ \citenamefont {Liu}}]{zhang2021insights}%
  \BibitemOpen
  \bibfield  {author} {\bibinfo {author} {\bibfnamefont {S.}~\bibnamefont
  {Zhang}}, \bibinfo {author} {\bibfnamefont {J.}~\bibnamefont {Liu}},\ and\
  \bibinfo {author} {\bibfnamefont {L.}~\bibnamefont {Liu}},\ }\bibfield
  {title} {\bibinfo {title} {Insights into the thermal conductivity of {MOF}-5
  from first principles},\ }\href {https://doi.org/10.1039/D1RA07022C}
  {\bibfield  {journal} {\bibinfo  {journal} {RSC Advances}\ }\textbf {\bibinfo
  {volume} {11}},\ \bibinfo {pages} {36928} (\bibinfo {year}
  {2021})}\BibitemShut {NoStop}%
\bibitem [{\citenamefont {Lamaire}\ \emph {et~al.}(2021)\citenamefont
  {Lamaire}, \citenamefont {Wieme}, \citenamefont {Hoffman},\ and\
  \citenamefont {Van~Speybroeck}}]{lamaire2021atomistic}%
  \BibitemOpen
  \bibfield  {author} {\bibinfo {author} {\bibfnamefont {A.}~\bibnamefont
  {Lamaire}}, \bibinfo {author} {\bibfnamefont {J.}~\bibnamefont {Wieme}},
  \bibinfo {author} {\bibfnamefont {A.~E.}\ \bibnamefont {Hoffman}},\ and\
  \bibinfo {author} {\bibfnamefont {V.}~\bibnamefont {Van~Speybroeck}},\
  }\bibfield  {title} {\bibinfo {title} {Atomistic insight in the flexibility
  and heat transport properties of the stimuli-responsive metal--organic
  framework {MIL-53 (Al)} for water-adsorption applications using molecular
  simulations},\ }\href {https://doi.org/10.1039/D0FD00025F} {\bibfield
  {journal} {\bibinfo  {journal} {Faraday Discussions}\ }\textbf {\bibinfo
  {volume} {225}},\ \bibinfo {pages} {301} (\bibinfo {year}
  {2021})}\BibitemShut {NoStop}%
\bibitem [{\citenamefont {Wieser}\ \emph {et~al.}(2021)\citenamefont {Wieser},
  \citenamefont {Kamencek}, \citenamefont {D{\"u}rholt}, \citenamefont
  {Schmid}, \citenamefont {Bedoya-Mart{\'\i}nez},\ and\ \citenamefont
  {Zojer}}]{wieser2021identifying}%
  \BibitemOpen
  \bibfield  {author} {\bibinfo {author} {\bibfnamefont {S.}~\bibnamefont
  {Wieser}}, \bibinfo {author} {\bibfnamefont {T.}~\bibnamefont {Kamencek}},
  \bibinfo {author} {\bibfnamefont {J.~P.}\ \bibnamefont {D{\"u}rholt}},
  \bibinfo {author} {\bibfnamefont {R.}~\bibnamefont {Schmid}}, \bibinfo
  {author} {\bibfnamefont {N.}~\bibnamefont {Bedoya-Mart{\'\i}nez}},\ and\
  \bibinfo {author} {\bibfnamefont {E.}~\bibnamefont {Zojer}},\ }\bibfield
  {title} {\bibinfo {title} {Identifying the bottleneck for heat transport in
  metal--organic frameworks},\ }\href {https://doi.org/10.1002/adts.202000211}
  {\bibfield  {journal} {\bibinfo  {journal} {Advanced Theory and Simulations}\
  }\textbf {\bibinfo {volume} {4}},\ \bibinfo {pages} {2000211} (\bibinfo
  {year} {2021})}\BibitemShut {NoStop}%
\bibitem [{\citenamefont {Zhou}\ \emph {et~al.}(2021)\citenamefont {Zhou},
  \citenamefont {Huang},\ and\ \citenamefont {Cao}}]{zhou2021vibrational}%
  \BibitemOpen
  \bibfield  {author} {\bibinfo {author} {\bibfnamefont {Y.}~\bibnamefont
  {Zhou}}, \bibinfo {author} {\bibfnamefont {B.}~\bibnamefont {Huang}},\ and\
  \bibinfo {author} {\bibfnamefont {B.-Y.}\ \bibnamefont {Cao}},\ }\bibfield
  {title} {\bibinfo {title} {Vibrational modes with long mean free path and
  large volumetric heat capacity drive higher thermal conductivity in amorphous
  zeolitic imidazolate framework-4},\ }\href
  {https://doi.org/10.1016/j.mtphys.2021.100516} {\bibfield  {journal}
  {\bibinfo  {journal} {Materials Today Physics}\ }\textbf {\bibinfo {volume}
  {21}},\ \bibinfo {pages} {100516} (\bibinfo {year} {2021})}\BibitemShut
  {NoStop}%
\bibitem [{\citenamefont {Zhou}\ \emph {et~al.}(2022)\citenamefont {Zhou},
  \citenamefont {Xu}, \citenamefont {Gao},\ and\ \citenamefont
  {Volz}}]{zhou2022origin}%
  \BibitemOpen
  \bibfield  {author} {\bibinfo {author} {\bibfnamefont {Y.}~\bibnamefont
  {Zhou}}, \bibinfo {author} {\bibfnamefont {Y.}~\bibnamefont {Xu}}, \bibinfo
  {author} {\bibfnamefont {Y.}~\bibnamefont {Gao}},\ and\ \bibinfo {author}
  {\bibfnamefont {S.}~\bibnamefont {Volz}},\ }\bibfield  {title} {\bibinfo
  {title} {Origin of the weakly temperature-dependent thermal conductivity in
  {ZIF-4 and ZIF-62}},\ }\href
  {https://doi.org/10.1103/PhysRevMaterials.6.015403} {\bibfield  {journal}
  {\bibinfo  {journal} {Physical Review Materials}\ }\textbf {\bibinfo {volume}
  {6}},\ \bibinfo {pages} {015403} (\bibinfo {year} {2022})}\BibitemShut
  {NoStop}%
\bibitem [{\citenamefont {Wieser}\ \emph {et~al.}(2022)\citenamefont {Wieser},
  \citenamefont {Kamencek}, \citenamefont {Schmid}, \citenamefont
  {Bedoya-Mart{\'\i}nez},\ and\ \citenamefont {Zojer}}]{wieser2022exploring}%
  \BibitemOpen
  \bibfield  {author} {\bibinfo {author} {\bibfnamefont {S.}~\bibnamefont
  {Wieser}}, \bibinfo {author} {\bibfnamefont {T.}~\bibnamefont {Kamencek}},
  \bibinfo {author} {\bibfnamefont {R.}~\bibnamefont {Schmid}}, \bibinfo
  {author} {\bibfnamefont {N.}~\bibnamefont {Bedoya-Mart{\'\i}nez}},\ and\
  \bibinfo {author} {\bibfnamefont {E.}~\bibnamefont {Zojer}},\ }\bibfield
  {title} {\bibinfo {title} {{Exploring the Impact of the Linker Length on Heat
  Transport in Metal--Organic Frameworks}},\ }\href
  {https://doi.org/10.3390/nano12132142} {\bibfield  {journal} {\bibinfo
  {journal} {Nanomaterials}\ }\textbf {\bibinfo {volume} {12}},\ \bibinfo
  {pages} {2142} (\bibinfo {year} {2022})}\BibitemShut {NoStop}%
\bibitem [{\citenamefont {Fan}\ \emph {et~al.}(2022{\natexlab{a}})\citenamefont
  {Fan}, \citenamefont {Yang},\ and\ \citenamefont {Zhou}}]{fan2022ultralong}%
  \BibitemOpen
  \bibfield  {author} {\bibinfo {author} {\bibfnamefont {H.}~\bibnamefont
  {Fan}}, \bibinfo {author} {\bibfnamefont {C.}~\bibnamefont {Yang}},\ and\
  \bibinfo {author} {\bibfnamefont {Y.}~\bibnamefont {Zhou}},\ }\bibfield
  {title} {\bibinfo {title} {Ultralong mean free path phonons in {HKUST-1} and
  their scattering by water adsorbates},\ }\href
  {https://doi.org/10.1103/PhysRevB.106.085417} {\bibfield  {journal} {\bibinfo
   {journal} {Physical Review B}\ }\textbf {\bibinfo {volume} {106}},\ \bibinfo
  {pages} {085417} (\bibinfo {year} {2022}{\natexlab{a}})}\BibitemShut
  {NoStop}%
\bibitem [{\citenamefont {Li}\ \emph {et~al.}(1999)\citenamefont {Li},
  \citenamefont {Eddaoudi}, \citenamefont {O'Keeffe},\ and\ \citenamefont
  {Yaghi}}]{li1999design}%
  \BibitemOpen
  \bibfield  {author} {\bibinfo {author} {\bibfnamefont {H.}~\bibnamefont
  {Li}}, \bibinfo {author} {\bibfnamefont {M.}~\bibnamefont {Eddaoudi}},
  \bibinfo {author} {\bibfnamefont {M.}~\bibnamefont {O'Keeffe}},\ and\
  \bibinfo {author} {\bibfnamefont {O.~M.}\ \bibnamefont {Yaghi}},\ }\bibfield
  {title} {\bibinfo {title} {Design and synthesis of an exceptionally stable
  and highly porous metal-organic framework},\ }\href
  {https://doi.org/10.1038/46248} {\bibfield  {journal} {\bibinfo  {journal}
  {Nature}\ }\textbf {\bibinfo {volume} {402}},\ \bibinfo {pages} {276}
  (\bibinfo {year} {1999})}\BibitemShut {NoStop}%
\bibitem [{\citenamefont {Huang}\ \emph {et~al.}(2006)\citenamefont {Huang},
  \citenamefont {Lin}, \citenamefont {Zhang},\ and\ \citenamefont
  {Chen}}]{huang2006ligand}%
  \BibitemOpen
  \bibfield  {author} {\bibinfo {author} {\bibfnamefont {X.-C.}\ \bibnamefont
  {Huang}}, \bibinfo {author} {\bibfnamefont {Y.-Y.}\ \bibnamefont {Lin}},
  \bibinfo {author} {\bibfnamefont {J.-P.}\ \bibnamefont {Zhang}},\ and\
  \bibinfo {author} {\bibfnamefont {X.-M.}\ \bibnamefont {Chen}},\ }\bibfield
  {title} {\bibinfo {title} {{Ligand-directed strategy for zeolite-type
  metal--organic frameworks: zinc (II) imidazolates with unusual zeolitic
  topologies}},\ }\href {https://doi.org/10.1002/anie.200503778} {\bibfield
  {journal} {\bibinfo  {journal} {Angewandte Chemie International Edition}\
  }\textbf {\bibinfo {volume} {45}},\ \bibinfo {pages} {1557} (\bibinfo {year}
  {2006})}\BibitemShut {NoStop}%
\bibitem [{\citenamefont {Chui}\ \emph {et~al.}(1999)\citenamefont {Chui},
  \citenamefont {Lo}, \citenamefont {Charmant}, \citenamefont {Orpen},\ and\
  \citenamefont {Williams}}]{chui1999chemically}%
  \BibitemOpen
  \bibfield  {author} {\bibinfo {author} {\bibfnamefont {S.~S.-Y.}\
  \bibnamefont {Chui}}, \bibinfo {author} {\bibfnamefont {S.~M.-F.}\
  \bibnamefont {Lo}}, \bibinfo {author} {\bibfnamefont {J.~P.}\ \bibnamefont
  {Charmant}}, \bibinfo {author} {\bibfnamefont {A.~G.}\ \bibnamefont
  {Orpen}},\ and\ \bibinfo {author} {\bibfnamefont {I.~D.}\ \bibnamefont
  {Williams}},\ }\bibfield  {title} {\bibinfo {title} {{A chemically
  functionalizable nanoporous material [Cu3 (TMA) 2 (H2O) 3] n}},\ }\href
  {https://doi.org/10.1126/science.283.5405.1148} {\bibfield  {journal}
  {\bibinfo  {journal} {Science}\ }\textbf {\bibinfo {volume} {283}},\ \bibinfo
  {pages} {1148} (\bibinfo {year} {1999})}\BibitemShut {NoStop}%
\bibitem [{\citenamefont {Green}(1954)}]{green1954markoff}%
  \BibitemOpen
  \bibfield  {author} {\bibinfo {author} {\bibfnamefont {M.~S.}\ \bibnamefont
  {Green}},\ }\bibfield  {title} {\bibinfo {title} {Markoff random processes
  and the statistical mechanics of time-dependent phenomena. ii. irreversible
  processes in fluids},\ }\href {https://doi.org/10.1063/1.1740082} {\bibfield
  {journal} {\bibinfo  {journal} {The Journal of Chemical Physics}\ }\textbf
  {\bibinfo {volume} {22}},\ \bibinfo {pages} {398} (\bibinfo {year}
  {1954})}\BibitemShut {NoStop}%
\bibitem [{\citenamefont {Kubo}(1957)}]{kubo1957statistical}%
  \BibitemOpen
  \bibfield  {author} {\bibinfo {author} {\bibfnamefont {R.}~\bibnamefont
  {Kubo}},\ }\bibfield  {title} {\bibinfo {title} {Statistical-mechanical
  theory of irreversible processes. i. general theory and simple applications
  to magnetic and conduction problems},\ }\href
  {https://doi.org/10.1143/JPSJ.12.570} {\bibfield  {journal} {\bibinfo
  {journal} {Journal of the Physical Society of Japan}\ }\textbf {\bibinfo
  {volume} {12}},\ \bibinfo {pages} {570} (\bibinfo {year} {1957})}\BibitemShut
  {NoStop}%
\bibitem [{\citenamefont {Stukowski}(2009)}]{stukowski2009visualization}%
  \BibitemOpen
  \bibfield  {author} {\bibinfo {author} {\bibfnamefont {A.}~\bibnamefont
  {Stukowski}},\ }\bibfield  {title} {\bibinfo {title} {{Visualization and
  analysis of atomistic simulation data with OVITO--the Open Visualization
  Tool}},\ }\href {https://doi.org/10.1088/0965-0393/18/1/015012} {\bibfield
  {journal} {\bibinfo  {journal} {Modelling and Simulation in Materials Science
  and Engineering}\ }\textbf {\bibinfo {volume} {18}},\ \bibinfo {pages}
  {015012} (\bibinfo {year} {2009})}\BibitemShut {NoStop}%
\bibitem [{\citenamefont {Thompson}\ \emph {et~al.}(2022)\citenamefont
  {Thompson}, \citenamefont {Aktulga}, \citenamefont {Berger}, \citenamefont
  {Bolintineanu}, \citenamefont {Brown}, \citenamefont {Crozier}, \citenamefont
  {in't Veld}, \citenamefont {Kohlmeyer}, \citenamefont {Moore}, \citenamefont
  {Nguyen} \emph {et~al.}}]{thompson2022lammps}%
  \BibitemOpen
  \bibfield  {author} {\bibinfo {author} {\bibfnamefont {A.~P.}\ \bibnamefont
  {Thompson}}, \bibinfo {author} {\bibfnamefont {H.~M.}\ \bibnamefont
  {Aktulga}}, \bibinfo {author} {\bibfnamefont {R.}~\bibnamefont {Berger}},
  \bibinfo {author} {\bibfnamefont {D.~S.}\ \bibnamefont {Bolintineanu}},
  \bibinfo {author} {\bibfnamefont {W.~M.}\ \bibnamefont {Brown}}, \bibinfo
  {author} {\bibfnamefont {P.~S.}\ \bibnamefont {Crozier}}, \bibinfo {author}
  {\bibfnamefont {P.~J.}\ \bibnamefont {in't Veld}}, \bibinfo {author}
  {\bibfnamefont {A.}~\bibnamefont {Kohlmeyer}}, \bibinfo {author}
  {\bibfnamefont {S.~G.}\ \bibnamefont {Moore}}, \bibinfo {author}
  {\bibfnamefont {T.~D.}\ \bibnamefont {Nguyen}}, \emph {et~al.},\ }\bibfield
  {title} {\bibinfo {title} {{LAMMPS-a flexible simulation tool for
  particle-based materials modeling at the atomic, meso, and continuum
  scales}},\ }\href {https://doi.org/10.1016/j.cpc.2021.108171} {\bibfield
  {journal} {\bibinfo  {journal} {Computer Physics Communications}\ }\textbf
  {\bibinfo {volume} {271}},\ \bibinfo {pages} {108171} (\bibinfo {year}
  {2022})}\BibitemShut {NoStop}%
\bibitem [{\citenamefont {Fan}\ \emph {et~al.}(2015)\citenamefont {Fan},
  \citenamefont {Pereira}, \citenamefont {Wang}, \citenamefont {Zheng},
  \citenamefont {Donadio},\ and\ \citenamefont {Harju}}]{fan2015force}%
  \BibitemOpen
  \bibfield  {author} {\bibinfo {author} {\bibfnamefont {Z.}~\bibnamefont
  {Fan}}, \bibinfo {author} {\bibfnamefont {L.~F.~C.}\ \bibnamefont {Pereira}},
  \bibinfo {author} {\bibfnamefont {H.-Q.}\ \bibnamefont {Wang}}, \bibinfo
  {author} {\bibfnamefont {J.-C.}\ \bibnamefont {Zheng}}, \bibinfo {author}
  {\bibfnamefont {D.}~\bibnamefont {Donadio}},\ and\ \bibinfo {author}
  {\bibfnamefont {A.}~\bibnamefont {Harju}},\ }\bibfield  {title} {\bibinfo
  {title} {{Force and heat current formulas for many-body potentials in
  molecular dynamics simulations with applications to thermal conductivity
  calculations}},\ }\href {https://doi.org/10.1103/PhysRevB.92.094301}
  {\bibfield  {journal} {\bibinfo  {journal} {Physical Review B}\ }\textbf
  {\bibinfo {volume} {92}},\ \bibinfo {pages} {094301} (\bibinfo {year}
  {2015})}\BibitemShut {NoStop}%
\bibitem [{\citenamefont {Tayfuroglu}\ \emph {et~al.}(2022)\citenamefont
  {Tayfuroglu}, \citenamefont {Kocak},\ and\ \citenamefont
  {Zorlu}}]{tayfuroglu2022neural}%
  \BibitemOpen
  \bibfield  {author} {\bibinfo {author} {\bibfnamefont {O.}~\bibnamefont
  {Tayfuroglu}}, \bibinfo {author} {\bibfnamefont {A.}~\bibnamefont {Kocak}},\
  and\ \bibinfo {author} {\bibfnamefont {Y.}~\bibnamefont {Zorlu}},\ }\bibfield
   {title} {\bibinfo {title} {{A neural network potential for the IRMOF series
  and its application for thermal and mechanical behaviors}},\ }\href
  {https://doi.org/10.1039/D1CP05973D} {\bibfield  {journal} {\bibinfo
  {journal} {Physical Chemistry Chemical Physics}\ }\textbf {\bibinfo {volume}
  {24}},\ \bibinfo {pages} {11882} (\bibinfo {year} {2022})}\BibitemShut
  {NoStop}%
\bibitem [{\citenamefont {Vandenhaute}\ \emph {et~al.}(2023)\citenamefont
  {Vandenhaute}, \citenamefont {Cools-Ceuppens}, \citenamefont {DeKeyser},
  \citenamefont {Verstraelen},\ and\ \citenamefont
  {Van~Speybroeck}}]{vandenhaute2023machine}%
  \BibitemOpen
  \bibfield  {author} {\bibinfo {author} {\bibfnamefont {S.}~\bibnamefont
  {Vandenhaute}}, \bibinfo {author} {\bibfnamefont {M.}~\bibnamefont
  {Cools-Ceuppens}}, \bibinfo {author} {\bibfnamefont {S.}~\bibnamefont
  {DeKeyser}}, \bibinfo {author} {\bibfnamefont {T.}~\bibnamefont
  {Verstraelen}},\ and\ \bibinfo {author} {\bibfnamefont {V.}~\bibnamefont
  {Van~Speybroeck}},\ }\bibfield  {title} {\bibinfo {title} {{Machine learning
  potentials for metal-organic frameworks using an incremental learning
  approach}},\ }\href {https://doi.org/10.1038/s41524-023-00969-x} {\bibfield
  {journal} {\bibinfo  {journal} {npj Computational Materials}\ }\textbf
  {\bibinfo {volume} {9}},\ \bibinfo {pages} {19} (\bibinfo {year}
  {2023})}\BibitemShut {NoStop}%
\bibitem [{\citenamefont {Achar}\ \emph {et~al.}(2022)\citenamefont {Achar},
  \citenamefont {Wardzala}, \citenamefont {Bernasconi}, \citenamefont {Zhang},\
  and\ \citenamefont {Johnson}}]{achar2022combined}%
  \BibitemOpen
  \bibfield  {author} {\bibinfo {author} {\bibfnamefont {S.~K.}\ \bibnamefont
  {Achar}}, \bibinfo {author} {\bibfnamefont {J.~J.}\ \bibnamefont {Wardzala}},
  \bibinfo {author} {\bibfnamefont {L.}~\bibnamefont {Bernasconi}}, \bibinfo
  {author} {\bibfnamefont {L.}~\bibnamefont {Zhang}},\ and\ \bibinfo {author}
  {\bibfnamefont {J.~K.}\ \bibnamefont {Johnson}},\ }\bibfield  {title}
  {\bibinfo {title} {Combined deep learning and classical potential approach
  for modeling diffusion in {UiO-66}},\ }\href
  {https://doi.org/10.1021/acs.jctc.2c00010} {\bibfield  {journal} {\bibinfo
  {journal} {Journal of Chemical Theory and Computation}\ }\textbf {\bibinfo
  {volume} {18}},\ \bibinfo {pages} {3593} (\bibinfo {year}
  {2022})}\BibitemShut {NoStop}%
\bibitem [{\citenamefont {Zheng}\ \emph {et~al.}(2023)\citenamefont {Zheng},
  \citenamefont {Oliveira}, \citenamefont {Neumann Barros~Ferreira},
  \citenamefont {Steiner}, \citenamefont {Hamann}, \citenamefont {Gu},\ and\
  \citenamefont {Luan}}]{zheng2023quantum}%
  \BibitemOpen
  \bibfield  {author} {\bibinfo {author} {\bibfnamefont {B.}~\bibnamefont
  {Zheng}}, \bibinfo {author} {\bibfnamefont {F.~L.}\ \bibnamefont {Oliveira}},
  \bibinfo {author} {\bibfnamefont {R.}~\bibnamefont {Neumann
  Barros~Ferreira}}, \bibinfo {author} {\bibfnamefont {M.}~\bibnamefont
  {Steiner}}, \bibinfo {author} {\bibfnamefont {H.}~\bibnamefont {Hamann}},
  \bibinfo {author} {\bibfnamefont {G.~X.}\ \bibnamefont {Gu}},\ and\ \bibinfo
  {author} {\bibfnamefont {B.}~\bibnamefont {Luan}},\ }\bibfield  {title}
  {\bibinfo {title} {{Quantum Informed Machine-Learning Potentials for
  Molecular Dynamics Simulations of CO2’s Chemisorption and Diffusion in
  Mg-MOF-74}},\ }\bibfield  {journal} {\bibinfo  {journal} {ACS Nano}\ }\href
  {https://doi.org/10.1021/acsnano.2c11102} {10.1021/acsnano.2c11102} (\bibinfo
  {year} {2023})\BibitemShut {NoStop}%
\bibitem [{\citenamefont {Fan}\ \emph {et~al.}(2021)\citenamefont {Fan},
  \citenamefont {Zeng}, \citenamefont {Zhang}, \citenamefont {Wang},
  \citenamefont {Song}, \citenamefont {Dong}, \citenamefont {Chen},\ and\
  \citenamefont {Ala-Nissila}}]{fan2021neuroevolution}%
  \BibitemOpen
  \bibfield  {author} {\bibinfo {author} {\bibfnamefont {Z.}~\bibnamefont
  {Fan}}, \bibinfo {author} {\bibfnamefont {Z.}~\bibnamefont {Zeng}}, \bibinfo
  {author} {\bibfnamefont {C.}~\bibnamefont {Zhang}}, \bibinfo {author}
  {\bibfnamefont {Y.}~\bibnamefont {Wang}}, \bibinfo {author} {\bibfnamefont
  {K.}~\bibnamefont {Song}}, \bibinfo {author} {\bibfnamefont {H.}~\bibnamefont
  {Dong}}, \bibinfo {author} {\bibfnamefont {Y.}~\bibnamefont {Chen}},\ and\
  \bibinfo {author} {\bibfnamefont {T.}~\bibnamefont {Ala-Nissila}},\
  }\bibfield  {title} {\bibinfo {title} {{Neuroevolution machine learning
  potentials: Combining high accuracy and low cost in atomistic simulations and
  application to heat transport}},\ }\href
  {https://doi.org/10.1103/PhysRevB.104.104309} {\bibfield  {journal} {\bibinfo
   {journal} {Physical Review B}\ }\textbf {\bibinfo {volume} {104}},\ \bibinfo
  {pages} {104309} (\bibinfo {year} {2021})}\BibitemShut {NoStop}%
\bibitem [{\citenamefont {Fan}(2022)}]{fan2022improving}%
  \BibitemOpen
  \bibfield  {author} {\bibinfo {author} {\bibfnamefont {Z.}~\bibnamefont
  {Fan}},\ }\bibfield  {title} {\bibinfo {title} {Improving the accuracy of the
  neuroevolution machine learning potential for multi-component systems},\
  }\href {https://doi.org/10.1088/1361-648X/ac462b} {\bibfield  {journal}
  {\bibinfo  {journal} {Journal of Physics: Condensed Matter}\ }\textbf
  {\bibinfo {volume} {34}},\ \bibinfo {pages} {125902} (\bibinfo {year}
  {2022})}\BibitemShut {NoStop}%
\bibitem [{\citenamefont {Fan}\ \emph {et~al.}(2022{\natexlab{b}})\citenamefont
  {Fan}, \citenamefont {Wang}, \citenamefont {Ying}, \citenamefont {Song},
  \citenamefont {Wang}, \citenamefont {Wang}, \citenamefont {Zeng},
  \citenamefont {Xu}, \citenamefont {Lindgren}, \citenamefont {Rahm},
  \citenamefont {Gabourie}, \citenamefont {Liu}, \citenamefont {Dong},
  \citenamefont {Wu}, \citenamefont {Chen}, \citenamefont {Zhong},
  \citenamefont {Sun}, \citenamefont {Erhart}, \citenamefont {Su},\ and\
  \citenamefont {Ala-Nissila}}]{Fan2022GPUMD}%
  \BibitemOpen
  \bibfield  {author} {\bibinfo {author} {\bibfnamefont {Z.}~\bibnamefont
  {Fan}}, \bibinfo {author} {\bibfnamefont {Y.}~\bibnamefont {Wang}}, \bibinfo
  {author} {\bibfnamefont {P.}~\bibnamefont {Ying}}, \bibinfo {author}
  {\bibfnamefont {K.}~\bibnamefont {Song}}, \bibinfo {author} {\bibfnamefont
  {J.}~\bibnamefont {Wang}}, \bibinfo {author} {\bibfnamefont {Y.}~\bibnamefont
  {Wang}}, \bibinfo {author} {\bibfnamefont {Z.}~\bibnamefont {Zeng}}, \bibinfo
  {author} {\bibfnamefont {K.}~\bibnamefont {Xu}}, \bibinfo {author}
  {\bibfnamefont {E.}~\bibnamefont {Lindgren}}, \bibinfo {author}
  {\bibfnamefont {J.~M.}\ \bibnamefont {Rahm}}, \bibinfo {author}
  {\bibfnamefont {A.~J.}\ \bibnamefont {Gabourie}}, \bibinfo {author}
  {\bibfnamefont {J.}~\bibnamefont {Liu}}, \bibinfo {author} {\bibfnamefont
  {H.}~\bibnamefont {Dong}}, \bibinfo {author} {\bibfnamefont {J.}~\bibnamefont
  {Wu}}, \bibinfo {author} {\bibfnamefont {Y.}~\bibnamefont {Chen}}, \bibinfo
  {author} {\bibfnamefont {Z.}~\bibnamefont {Zhong}}, \bibinfo {author}
  {\bibfnamefont {J.}~\bibnamefont {Sun}}, \bibinfo {author} {\bibfnamefont
  {P.}~\bibnamefont {Erhart}}, \bibinfo {author} {\bibfnamefont
  {Y.}~\bibnamefont {Su}},\ and\ \bibinfo {author} {\bibfnamefont
  {T.}~\bibnamefont {Ala-Nissila}},\ }\bibfield  {title} {\bibinfo {title}
  {{GPUMD}: A package for constructing accurate machine-learned potentials and
  performing highly efficient atomistic simulations},\ }\href
  {https://doi.org/10.1063/5.0106617} {\bibfield  {journal} {\bibinfo
  {journal} {The Journal of Chemical Physics}\ }\textbf {\bibinfo {volume}
  {157}},\ \bibinfo {pages} {114801} (\bibinfo {year}
  {2022}{\natexlab{b}})}\BibitemShut {NoStop}%
\bibitem [{\citenamefont {Fan}\ \emph {et~al.}(2017)\citenamefont {Fan},
  \citenamefont {Chen}, \citenamefont {Vierimaa},\ and\ \citenamefont
  {Harju}}]{Fan2017CPC}%
  \BibitemOpen
  \bibfield  {author} {\bibinfo {author} {\bibfnamefont {Z.}~\bibnamefont
  {Fan}}, \bibinfo {author} {\bibfnamefont {W.}~\bibnamefont {Chen}}, \bibinfo
  {author} {\bibfnamefont {V.}~\bibnamefont {Vierimaa}},\ and\ \bibinfo
  {author} {\bibfnamefont {A.}~\bibnamefont {Harju}},\ }\bibfield  {title}
  {\bibinfo {title} {{Efficient molecular dynamics simulations with many-body
  potentials on graphics processing units}},\ }\href
  {https://doi.org/10.1016/j.cpc.2017.05.003} {\bibfield  {journal} {\bibinfo
  {journal} {Computer Physics Communications}\ }\textbf {\bibinfo {volume}
  {218}},\ \bibinfo {pages} {10} (\bibinfo {year} {2017})}\BibitemShut
  {NoStop}%
\bibitem [{\citenamefont {Fan}\ \emph {et~al.}(2019)\citenamefont {Fan},
  \citenamefont {Dong}, \citenamefont {Harju},\ and\ \citenamefont
  {Ala-Nissila}}]{Fan2019PRB}%
  \BibitemOpen
  \bibfield  {author} {\bibinfo {author} {\bibfnamefont {Z.}~\bibnamefont
  {Fan}}, \bibinfo {author} {\bibfnamefont {H.}~\bibnamefont {Dong}}, \bibinfo
  {author} {\bibfnamefont {A.}~\bibnamefont {Harju}},\ and\ \bibinfo {author}
  {\bibfnamefont {T.}~\bibnamefont {Ala-Nissila}},\ }\bibfield  {title}
  {\bibinfo {title} {{Homogeneous nonequilibrium molecular dynamics method for
  heat transport and spectral decomposition with many-body potentials}},\
  }\href {https://doi.org/10.1103/PhysRevB.99.064308} {\bibfield  {journal}
  {\bibinfo  {journal} {Physical Review B}\ }\textbf {\bibinfo {volume} {99}},\
  \bibinfo {pages} {064308} (\bibinfo {year} {2019})}\BibitemShut {NoStop}%
\bibitem [{\citenamefont {Addicoat}\ \emph {et~al.}(2014)\citenamefont
  {Addicoat}, \citenamefont {Vankova}, \citenamefont {Akter},\ and\
  \citenamefont {Heine}}]{addicoat2014extension}%
  \BibitemOpen
  \bibfield  {author} {\bibinfo {author} {\bibfnamefont {M.~A.}\ \bibnamefont
  {Addicoat}}, \bibinfo {author} {\bibfnamefont {N.}~\bibnamefont {Vankova}},
  \bibinfo {author} {\bibfnamefont {I.~F.}\ \bibnamefont {Akter}},\ and\
  \bibinfo {author} {\bibfnamefont {T.}~\bibnamefont {Heine}},\ }\bibfield
  {title} {\bibinfo {title} {{Extension of the universal force field to
  metal--organic frameworks}},\ }\href {https://doi.org/10.1021/ct400952t}
  {\bibfield  {journal} {\bibinfo  {journal} {Journal of Chemical Theory and
  Computation}\ }\textbf {\bibinfo {volume} {10}},\ \bibinfo {pages} {880}
  (\bibinfo {year} {2014})}\BibitemShut {NoStop}%
\bibitem [{\citenamefont {Boyd}\ \emph {et~al.}(2017)\citenamefont {Boyd},
  \citenamefont {Moosavi}, \citenamefont {Witman},\ and\ \citenamefont
  {Smit}}]{boyd2017force}%
  \BibitemOpen
  \bibfield  {author} {\bibinfo {author} {\bibfnamefont {P.~G.}\ \bibnamefont
  {Boyd}}, \bibinfo {author} {\bibfnamefont {S.~M.}\ \bibnamefont {Moosavi}},
  \bibinfo {author} {\bibfnamefont {M.}~\bibnamefont {Witman}},\ and\ \bibinfo
  {author} {\bibfnamefont {B.}~\bibnamefont {Smit}},\ }\bibfield  {title}
  {\bibinfo {title} {{Force-field prediction of materials properties in
  metal-organic frameworks}},\ }\href
  {https://doi.org/10.1021/acs.jpclett.6b02532} {\bibfield  {journal} {\bibinfo
   {journal} {The journal of physical chemistry letters}\ }\textbf {\bibinfo
  {volume} {8}},\ \bibinfo {pages} {357} (\bibinfo {year} {2017})}\BibitemShut
  {NoStop}%
\bibitem [{\citenamefont {Pallach}\ \emph {et~al.}(2021)\citenamefont
  {Pallach}, \citenamefont {Keupp}, \citenamefont {Terlinden}, \citenamefont
  {Frentzel-Beyme}, \citenamefont {Klo{\ss}}, \citenamefont {Machalica},
  \citenamefont {Kotschy}, \citenamefont {Vasa}, \citenamefont {Chater},
  \citenamefont {Sternemann} \emph {et~al.}}]{pallach2021frustrated}%
  \BibitemOpen
  \bibfield  {author} {\bibinfo {author} {\bibfnamefont {R.}~\bibnamefont
  {Pallach}}, \bibinfo {author} {\bibfnamefont {J.}~\bibnamefont {Keupp}},
  \bibinfo {author} {\bibfnamefont {K.}~\bibnamefont {Terlinden}}, \bibinfo
  {author} {\bibfnamefont {L.}~\bibnamefont {Frentzel-Beyme}}, \bibinfo
  {author} {\bibfnamefont {M.}~\bibnamefont {Klo{\ss}}}, \bibinfo {author}
  {\bibfnamefont {A.}~\bibnamefont {Machalica}}, \bibinfo {author}
  {\bibfnamefont {J.}~\bibnamefont {Kotschy}}, \bibinfo {author} {\bibfnamefont
  {S.~K.}\ \bibnamefont {Vasa}}, \bibinfo {author} {\bibfnamefont {P.~A.}\
  \bibnamefont {Chater}}, \bibinfo {author} {\bibfnamefont {C.}~\bibnamefont
  {Sternemann}}, \emph {et~al.},\ }\bibfield  {title} {\bibinfo {title}
  {{Frustrated flexibility in metal-organic frameworks}},\ }\href
  {https://doi.org/10.1038/s41467-021-24188-4} {\bibfield  {journal} {\bibinfo
  {journal} {Nature communications}\ }\textbf {\bibinfo {volume} {12}},\
  \bibinfo {pages} {4097} (\bibinfo {year} {2021})}\BibitemShut {NoStop}%
\bibitem [{\citenamefont {Bureekaew}\ \emph {et~al.}(2013)\citenamefont
  {Bureekaew}, \citenamefont {Amirjalayer}, \citenamefont {Tafipolsky},
  \citenamefont {Spickermann}, \citenamefont {Roy},\ and\ \citenamefont
  {Schmid}}]{bureekaew2013mof}%
  \BibitemOpen
  \bibfield  {author} {\bibinfo {author} {\bibfnamefont {S.}~\bibnamefont
  {Bureekaew}}, \bibinfo {author} {\bibfnamefont {S.}~\bibnamefont
  {Amirjalayer}}, \bibinfo {author} {\bibfnamefont {M.}~\bibnamefont
  {Tafipolsky}}, \bibinfo {author} {\bibfnamefont {C.}~\bibnamefont
  {Spickermann}}, \bibinfo {author} {\bibfnamefont {T.~K.}\ \bibnamefont
  {Roy}},\ and\ \bibinfo {author} {\bibfnamefont {R.}~\bibnamefont {Schmid}},\
  }\bibfield  {title} {\bibinfo {title} {{MOF-FF--A flexible first-principles
  derived force field for metal-organic frameworks}},\ }\href
  {https://doi.org/10.1002/pssb.201248460} {\bibfield  {journal} {\bibinfo
  {journal} {physica status solidi (b)}\ }\textbf {\bibinfo {volume} {250}},\
  \bibinfo {pages} {1128} (\bibinfo {year} {2013})}\BibitemShut {NoStop}%
\bibitem [{\citenamefont {D{\"u}rholt}\ \emph {et~al.}(2019)\citenamefont
  {D{\"u}rholt}, \citenamefont {Fraux}, \citenamefont {Coudert},\ and\
  \citenamefont {Schmid}}]{durholt2019ab}%
  \BibitemOpen
  \bibfield  {author} {\bibinfo {author} {\bibfnamefont {J.~P.}\ \bibnamefont
  {D{\"u}rholt}}, \bibinfo {author} {\bibfnamefont {G.}~\bibnamefont {Fraux}},
  \bibinfo {author} {\bibfnamefont {F.-X.}\ \bibnamefont {Coudert}},\ and\
  \bibinfo {author} {\bibfnamefont {R.}~\bibnamefont {Schmid}},\ }\bibfield
  {title} {\bibinfo {title} {{Ab initio derived force fields for zeolitic
  imidazolate frameworks: MOF-FF for ZIFs}},\ }\href
  {https://doi.org/10.1021/acs.jctc.8b01041} {\bibfield  {journal} {\bibinfo
  {journal} {Journal of Chemical Theory and Computation}\ }\textbf {\bibinfo
  {volume} {15}},\ \bibinfo {pages} {2420} (\bibinfo {year}
  {2019})}\BibitemShut {NoStop}%
\bibitem [{\citenamefont {Monti}\ \emph {et~al.}(2013)\citenamefont {Monti},
  \citenamefont {Li},\ and\ \citenamefont {Carravetta}}]{monti2013reactive}%
  \BibitemOpen
  \bibfield  {author} {\bibinfo {author} {\bibfnamefont {S.}~\bibnamefont
  {Monti}}, \bibinfo {author} {\bibfnamefont {C.}~\bibnamefont {Li}},\ and\
  \bibinfo {author} {\bibfnamefont {V.}~\bibnamefont {Carravetta}},\ }\bibfield
   {title} {\bibinfo {title} {{Reactive dynamics simulation of monolayer and
  multilayer adsorption of glycine on Cu (110)}},\ }\href
  {https://doi.org/10.1021/jp312828d} {\bibfield  {journal} {\bibinfo
  {journal} {The Journal of Physical Chemistry C}\ }\textbf {\bibinfo {volume}
  {117}},\ \bibinfo {pages} {5221} (\bibinfo {year} {2013})}\BibitemShut
  {NoStop}%
\bibitem [{\citenamefont {Han}\ \emph {et~al.}(2010)\citenamefont {Han},
  \citenamefont {Choi},\ and\ \citenamefont {Van~Duin}}]{han2010molecular}%
  \BibitemOpen
  \bibfield  {author} {\bibinfo {author} {\bibfnamefont {S.~S.}\ \bibnamefont
  {Han}}, \bibinfo {author} {\bibfnamefont {S.-H.}\ \bibnamefont {Choi}},\ and\
  \bibinfo {author} {\bibfnamefont {A.~C.}\ \bibnamefont {Van~Duin}},\
  }\bibfield  {title} {\bibinfo {title} {{Molecular dynamics simulations of
  stability of metal--organic frameworks against H 2 O using the ReaxFF
  reactive force field}},\ }\href {https://doi.org/10.1039/C0CC01132K}
  {\bibfield  {journal} {\bibinfo  {journal} {Chemical communications}\
  }\textbf {\bibinfo {volume} {46}},\ \bibinfo {pages} {5713} (\bibinfo {year}
  {2010})}\BibitemShut {NoStop}%
\bibitem [{\citenamefont {Yang}\ \emph {et~al.}(2018)\citenamefont {Yang},
  \citenamefont {Shin}, \citenamefont {Li}, \citenamefont {Bennett},
  \citenamefont {Van~Duin},\ and\ \citenamefont {Mauro}}]{yang2018enabling}%
  \BibitemOpen
  \bibfield  {author} {\bibinfo {author} {\bibfnamefont {Y.}~\bibnamefont
  {Yang}}, \bibinfo {author} {\bibfnamefont {Y.~K.}\ \bibnamefont {Shin}},
  \bibinfo {author} {\bibfnamefont {S.}~\bibnamefont {Li}}, \bibinfo {author}
  {\bibfnamefont {T.~D.}\ \bibnamefont {Bennett}}, \bibinfo {author}
  {\bibfnamefont {A.~C.}\ \bibnamefont {Van~Duin}},\ and\ \bibinfo {author}
  {\bibfnamefont {J.~C.}\ \bibnamefont {Mauro}},\ }\bibfield  {title} {\bibinfo
  {title} {{Enabling computational design of ZIFs using ReaxFF}},\ }\href
  {https://doi.org/10.1021/acs.jpcb.8b08094} {\bibfield  {journal} {\bibinfo
  {journal} {The Journal of Physical Chemistry B}\ }\textbf {\bibinfo {volume}
  {122}},\ \bibinfo {pages} {9616} (\bibinfo {year} {2018})}\BibitemShut
  {NoStop}%
\bibitem [{\citenamefont {Van~Duin}\ \emph {et~al.}(2001)\citenamefont
  {Van~Duin}, \citenamefont {Dasgupta}, \citenamefont {Lorant},\ and\
  \citenamefont {Goddard}}]{van2001reaxff}%
  \BibitemOpen
  \bibfield  {author} {\bibinfo {author} {\bibfnamefont {A.~C.}\ \bibnamefont
  {Van~Duin}}, \bibinfo {author} {\bibfnamefont {S.}~\bibnamefont {Dasgupta}},
  \bibinfo {author} {\bibfnamefont {F.}~\bibnamefont {Lorant}},\ and\ \bibinfo
  {author} {\bibfnamefont {W.~A.}\ \bibnamefont {Goddard}},\ }\bibfield
  {title} {\bibinfo {title} {{ReaxFF: a reactive force field for
  hydrocarbons}},\ }\href {https://doi.org/10.1021/jp004368u} {\bibfield
  {journal} {\bibinfo  {journal} {The Journal of Physical Chemistry A}\
  }\textbf {\bibinfo {volume} {105}},\ \bibinfo {pages} {9396} (\bibinfo {year}
  {2001})}\BibitemShut {NoStop}%
\bibitem [{\citenamefont {Banlusan}\ \emph {et~al.}(2015)\citenamefont
  {Banlusan}, \citenamefont {Antillon},\ and\ \citenamefont
  {Strachan}}]{banlusan2015mechanisms}%
  \BibitemOpen
  \bibfield  {author} {\bibinfo {author} {\bibfnamefont {K.}~\bibnamefont
  {Banlusan}}, \bibinfo {author} {\bibfnamefont {E.}~\bibnamefont {Antillon}},\
  and\ \bibinfo {author} {\bibfnamefont {A.}~\bibnamefont {Strachan}},\
  }\bibfield  {title} {\bibinfo {title} {{Mechanisms of Plastic Deformation of
  Metal--Organic Framework-5}},\ }\href
  {https://doi.org/10.1021/acs.jpcc.5b05446} {\bibfield  {journal} {\bibinfo
  {journal} {The Journal of Physical Chemistry C}\ }\textbf {\bibinfo {volume}
  {119}},\ \bibinfo {pages} {25845} (\bibinfo {year} {2015})}\BibitemShut
  {NoStop}%
\bibitem [{\citenamefont {Islamov}\ \emph {et~al.}(2023)\citenamefont
  {Islamov}, \citenamefont {Babaei}, \citenamefont {Anderson}, \citenamefont
  {Sezginel}, \citenamefont {Long}, \citenamefont {McGaughey}, \citenamefont
  {Gomez-Gualdron},\ and\ \citenamefont {Wilmer}}]{islamov2023high}%
  \BibitemOpen
  \bibfield  {author} {\bibinfo {author} {\bibfnamefont {M.}~\bibnamefont
  {Islamov}}, \bibinfo {author} {\bibfnamefont {H.}~\bibnamefont {Babaei}},
  \bibinfo {author} {\bibfnamefont {R.}~\bibnamefont {Anderson}}, \bibinfo
  {author} {\bibfnamefont {K.~B.}\ \bibnamefont {Sezginel}}, \bibinfo {author}
  {\bibfnamefont {J.~R.}\ \bibnamefont {Long}}, \bibinfo {author}
  {\bibfnamefont {A.~J.}\ \bibnamefont {McGaughey}}, \bibinfo {author}
  {\bibfnamefont {D.~A.}\ \bibnamefont {Gomez-Gualdron}},\ and\ \bibinfo
  {author} {\bibfnamefont {C.~E.}\ \bibnamefont {Wilmer}},\ }\bibfield  {title}
  {\bibinfo {title} {{High-throughput screening of hypothetical metal-organic
  frameworks for thermal conductivity}},\ }\href
  {https://doi.org/10.1038/s41524-022-00961-x} {\bibfield  {journal} {\bibinfo
  {journal} {npj Computational Materials}\ }\textbf {\bibinfo {volume} {9}},\
  \bibinfo {pages} {11} (\bibinfo {year} {2023})}\BibitemShut {NoStop}%
\bibitem [{\citenamefont {Evans}\ \emph {et~al.}(2019)\citenamefont {Evans},
  \citenamefont {D{\"u}rholt}, \citenamefont {Kaskel},\ and\ \citenamefont
  {Schmid}}]{evans2019assessing}%
  \BibitemOpen
  \bibfield  {author} {\bibinfo {author} {\bibfnamefont {J.~D.}\ \bibnamefont
  {Evans}}, \bibinfo {author} {\bibfnamefont {J.~P.}\ \bibnamefont
  {D{\"u}rholt}}, \bibinfo {author} {\bibfnamefont {S.}~\bibnamefont
  {Kaskel}},\ and\ \bibinfo {author} {\bibfnamefont {R.}~\bibnamefont
  {Schmid}},\ }\bibfield  {title} {\bibinfo {title} {{Assessing negative
  thermal expansion in mesoporous metal--organic frameworks by molecular
  simulation}},\ }\href {https://doi.org/10.1039/C9TA06644F} {\bibfield
  {journal} {\bibinfo  {journal} {Journal of Materials Chemistry A}\ }\textbf
  {\bibinfo {volume} {7}},\ \bibinfo {pages} {24019} (\bibinfo {year}
  {2019})}\BibitemShut {NoStop}%
\bibitem [{\citenamefont {Wang}\ \emph
  {et~al.}(2023{\natexlab{a}})\citenamefont {Wang}, \citenamefont {Ying},\ and\
  \citenamefont {Zhang}}]{wang2023effects}%
  \BibitemOpen
  \bibfield  {author} {\bibinfo {author} {\bibfnamefont {B.}~\bibnamefont
  {Wang}}, \bibinfo {author} {\bibfnamefont {P.}~\bibnamefont {Ying}},\ and\
  \bibinfo {author} {\bibfnamefont {J.}~\bibnamefont {Zhang}},\ }\bibfield
  {title} {\bibinfo {title} {{Effects of Missing Linker Defects on the Elastic
  Properties and Mechanical Stability of the Metal--Organic Framework
  HKUST-1}},\ }\href {https://doi.org/10.1021/acs.jpcc.2c06954} {\bibfield
  {journal} {\bibinfo  {journal} {The Journal of Physical Chemistry C}\
  }\textbf {\bibinfo {volume} {127}},\ \bibinfo {pages} {2533} (\bibinfo {year}
  {2023}{\natexlab{a}})}\BibitemShut {NoStop}%
\bibitem [{\citenamefont {Castel}\ and\ \citenamefont
  {Coudert}(2022)}]{castel2022atomistic}%
  \BibitemOpen
  \bibfield  {author} {\bibinfo {author} {\bibfnamefont {N.}~\bibnamefont
  {Castel}}\ and\ \bibinfo {author} {\bibfnamefont {F.-X.}\ \bibnamefont
  {Coudert}},\ }\bibfield  {title} {\bibinfo {title} {{Atomistic Models of
  Amorphous Metal--Organic Frameworks}},\ }\href
  {https://doi.org/10.1021/acs.jpcc.2c01091} {\bibfield  {journal} {\bibinfo
  {journal} {The Journal of Physical Chemistry C}\ }\textbf {\bibinfo {volume}
  {126}},\ \bibinfo {pages} {6905} (\bibinfo {year} {2022})}\BibitemShut
  {NoStop}%
\bibitem [{\citenamefont {Behler}\ and\ \citenamefont
  {Parrinello}(2007)}]{behler2007prl}%
  \BibitemOpen
  \bibfield  {author} {\bibinfo {author} {\bibfnamefont {J.}~\bibnamefont
  {Behler}}\ and\ \bibinfo {author} {\bibfnamefont {M.}~\bibnamefont
  {Parrinello}},\ }\bibfield  {title} {\bibinfo {title} {{Generalized
  neural-network representation of high-dimensional potential-energy
  surfaces}},\ }\href {https://doi.org/10.1103/PhysRevLett.98.146401}
  {\bibfield  {journal} {\bibinfo  {journal} {Physical Review Letters}\
  }\textbf {\bibinfo {volume} {98}},\ \bibinfo {pages} {146401} (\bibinfo
  {year} {2007})}\BibitemShut {NoStop}%
\bibitem [{\citenamefont {Drautz}(2019)}]{drautz2019atomic}%
  \BibitemOpen
  \bibfield  {author} {\bibinfo {author} {\bibfnamefont {R.}~\bibnamefont
  {Drautz}},\ }\bibfield  {title} {\bibinfo {title} {{Atomic cluster expansion
  for accurate and transferable interatomic potentials}},\ }\href
  {https://doi.org/10.1103/PhysRevB.99.014104} {\bibfield  {journal} {\bibinfo
  {journal} {Physical Review B}\ }\textbf {\bibinfo {volume} {99}},\ \bibinfo
  {pages} {014104} (\bibinfo {year} {2019})}\BibitemShut {NoStop}%
\bibitem [{\citenamefont {Schaul}\ \emph {et~al.}(2011)\citenamefont {Schaul},
  \citenamefont {Glasmachers},\ and\ \citenamefont {Schmidhuber}}]{Schaul2011}%
  \BibitemOpen
  \bibfield  {author} {\bibinfo {author} {\bibfnamefont {T.}~\bibnamefont
  {Schaul}}, \bibinfo {author} {\bibfnamefont {T.}~\bibnamefont
  {Glasmachers}},\ and\ \bibinfo {author} {\bibfnamefont {J.}~\bibnamefont
  {Schmidhuber}},\ }\bibfield  {title} {\bibinfo {title} {{High Dimensions and
  Heavy Tails for Natural Evolution Strategies}},\ }in\ \href
  {https://doi.org/10.1145/2001576.2001692} {\emph {\bibinfo {booktitle}
  {Proceedings of the 13th Annual Conference on Genetic and Evolutionary
  Computation}}},\ \bibinfo {series and number} {GECCO '11}\ (\bibinfo
  {publisher} {Association for Computing Machinery},\ \bibinfo {address} {New
  York, NY, USA},\ \bibinfo {year} {2011})\ pp.\ \bibinfo {pages}
  {845--852}\BibitemShut {NoStop}%
\bibitem [{\citenamefont {Wang}\ \emph
  {et~al.}(2023{\natexlab{b}})\citenamefont {Wang}, \citenamefont {Fan},
  \citenamefont {Qian}, \citenamefont {Caro},\ and\ \citenamefont
  {Ala-Nissila}}]{wang2022quantum}%
  \BibitemOpen
  \bibfield  {author} {\bibinfo {author} {\bibfnamefont {Y.}~\bibnamefont
  {Wang}}, \bibinfo {author} {\bibfnamefont {Z.}~\bibnamefont {Fan}}, \bibinfo
  {author} {\bibfnamefont {P.}~\bibnamefont {Qian}}, \bibinfo {author}
  {\bibfnamefont {M.~A.}\ \bibnamefont {Caro}},\ and\ \bibinfo {author}
  {\bibfnamefont {T.}~\bibnamefont {Ala-Nissila}},\ }\bibfield  {title}
  {\bibinfo {title} {Quantum-corrected thickness-dependent thermal conductivity
  in amorphous silicon predicted by machine learning molecular dynamics
  simulations},\ }\href {https://doi.org/10.1103/PhysRevB.107.054303}
  {\bibfield  {journal} {\bibinfo  {journal} {Physical Review B}\ }\textbf
  {\bibinfo {volume} {107}},\ \bibinfo {pages} {054303} (\bibinfo {year}
  {2023}{\natexlab{b}})}\BibitemShut {NoStop}%
\bibitem [{\citenamefont {Ying}\ \emph
  {et~al.}(2023{\natexlab{a}})\citenamefont {Ying}, \citenamefont {Liang},
  \citenamefont {Xu}, \citenamefont {Xu}, \citenamefont {Fan}, \citenamefont
  {Ala-Nissila},\ and\ \citenamefont {Zhong}}]{ying2023variable}%
  \BibitemOpen
  \bibfield  {author} {\bibinfo {author} {\bibfnamefont {P.}~\bibnamefont
  {Ying}}, \bibinfo {author} {\bibfnamefont {T.}~\bibnamefont {Liang}},
  \bibinfo {author} {\bibfnamefont {K.}~\bibnamefont {Xu}}, \bibinfo {author}
  {\bibfnamefont {J.}~\bibnamefont {Xu}}, \bibinfo {author} {\bibfnamefont
  {Z.}~\bibnamefont {Fan}}, \bibinfo {author} {\bibfnamefont {T.}~\bibnamefont
  {Ala-Nissila}},\ and\ \bibinfo {author} {\bibfnamefont {Z.}~\bibnamefont
  {Zhong}},\ }\bibfield  {title} {\bibinfo {title} {{Variable thermal transport
  in black, blue, and violet phosphorene from extensive atomistic simulations
  with a neuroevolution potential}},\ }\href
  {https://doi.org/10.1016/j.ijheatmasstransfer.2022.123681} {\bibfield
  {journal} {\bibinfo  {journal} {International Journal of Heat and Mass
  Transfer}\ }\textbf {\bibinfo {volume} {202}},\ \bibinfo {pages} {123681}
  (\bibinfo {year} {2023}{\natexlab{a}})}\BibitemShut {NoStop}%
\bibitem [{\citenamefont {Ying}\ \emph
  {et~al.}(2023{\natexlab{b}})\citenamefont {Ying}, \citenamefont {Dong},
  \citenamefont {Liang}, \citenamefont {Fan}, \citenamefont {Zhong},\ and\
  \citenamefont {Zhang}}]{ying2023atomistic}%
  \BibitemOpen
  \bibfield  {author} {\bibinfo {author} {\bibfnamefont {P.}~\bibnamefont
  {Ying}}, \bibinfo {author} {\bibfnamefont {H.}~\bibnamefont {Dong}}, \bibinfo
  {author} {\bibfnamefont {T.}~\bibnamefont {Liang}}, \bibinfo {author}
  {\bibfnamefont {Z.}~\bibnamefont {Fan}}, \bibinfo {author} {\bibfnamefont
  {Z.}~\bibnamefont {Zhong}},\ and\ \bibinfo {author} {\bibfnamefont
  {J.}~\bibnamefont {Zhang}},\ }\bibfield  {title} {\bibinfo {title}
  {{Atomistic insights into the mechanical anisotropy and fragility of
  monolayer fullerene networks using quantum mechanical calculations and
  machine-learning molecular dynamics simulations}},\ }\href
  {https://doi.org/10.1016/j.eml.2022.101929} {\bibfield  {journal} {\bibinfo
  {journal} {Extreme Mechanics Letters}\ }\textbf {\bibinfo {volume} {58}},\
  \bibinfo {pages} {101929} (\bibinfo {year} {2023}{\natexlab{b}})}\BibitemShut
  {NoStop}%
\bibitem [{\citenamefont {Dong}\ \emph {et~al.}(2023)\citenamefont {Dong},
  \citenamefont {Cao}, \citenamefont {Ying}, \citenamefont {Fan}, \citenamefont
  {Qian},\ and\ \citenamefont {Su}}]{dong2022anisotropic}%
  \BibitemOpen
  \bibfield  {author} {\bibinfo {author} {\bibfnamefont {H.}~\bibnamefont
  {Dong}}, \bibinfo {author} {\bibfnamefont {C.}~\bibnamefont {Cao}}, \bibinfo
  {author} {\bibfnamefont {P.}~\bibnamefont {Ying}}, \bibinfo {author}
  {\bibfnamefont {Z.}~\bibnamefont {Fan}}, \bibinfo {author} {\bibfnamefont
  {P.}~\bibnamefont {Qian}},\ and\ \bibinfo {author} {\bibfnamefont
  {Y.}~\bibnamefont {Su}},\ }\bibfield  {title} {\bibinfo {title} {{Anisotropic
  and high thermal conductivity in monolayer quasi-hexagonal fullerene: A
  comparative study against bulk phase fullerene}},\ }\href
  {https://doi.org/https://doi.org/10.1016/j.ijheatmasstransfer.2023.123943}
  {\bibfield  {journal} {\bibinfo  {journal} {International Journal of Heat and
  Mass Transfer}\ }\textbf {\bibinfo {volume} {206}},\ \bibinfo {pages}
  {123943} (\bibinfo {year} {2023})}\BibitemShut {NoStop}%
\bibitem [{\citenamefont {Eckhoff}\ and\ \citenamefont
  {Behler}(2019)}]{eckhoff2019molecular}%
  \BibitemOpen
  \bibfield  {author} {\bibinfo {author} {\bibfnamefont {M.}~\bibnamefont
  {Eckhoff}}\ and\ \bibinfo {author} {\bibfnamefont {J.}~\bibnamefont
  {Behler}},\ }\bibfield  {title} {\bibinfo {title} {{From molecular fragments
  to the bulk: development of a neural network potential for MOF-5}},\ }\href
  {https://doi.org/10.1021/acs.jctc.8b01288} {\bibfield  {journal} {\bibinfo
  {journal} {Journal of Chemical Theory and Computation}\ }\textbf {\bibinfo
  {volume} {15}},\ \bibinfo {pages} {3793} (\bibinfo {year}
  {2019})}\BibitemShut {NoStop}%
\bibitem [{\citenamefont {Perdew}\ \emph {et~al.}(1996)\citenamefont {Perdew},
  \citenamefont {Burke},\ and\ \citenamefont {Ernzerhof}}]{Perdew1996PRL}%
  \BibitemOpen
  \bibfield  {author} {\bibinfo {author} {\bibfnamefont {J.~P.}\ \bibnamefont
  {Perdew}}, \bibinfo {author} {\bibfnamefont {K.}~\bibnamefont {Burke}},\ and\
  \bibinfo {author} {\bibfnamefont {M.}~\bibnamefont {Ernzerhof}},\ }\bibfield
  {title} {\bibinfo {title} {{Generalized gradient approximation made
  simple}},\ }\href {https://doi.org/10.1103/PhysRevLett.77.3865} {\bibfield
  {journal} {\bibinfo  {journal} {Physical Review Letters}\ }\textbf {\bibinfo
  {volume} {77}},\ \bibinfo {pages} {3865} (\bibinfo {year}
  {1996})}\BibitemShut {NoStop}%
\bibitem [{\citenamefont {Bl{\"o}chl}(1994)}]{Blchl1994PRB}%
  \BibitemOpen
  \bibfield  {author} {\bibinfo {author} {\bibfnamefont {P.~E.}\ \bibnamefont
  {Bl{\"o}chl}},\ }\bibfield  {title} {\bibinfo {title} {{Projector
  augmented-wave method}},\ }\href {https://doi.org/10.1103/PhysRevB.50.17953}
  {\bibfield  {journal} {\bibinfo  {journal} {Physical Review B}\ }\textbf
  {\bibinfo {volume} {50}},\ \bibinfo {pages} {17953} (\bibinfo {year}
  {1994})}\BibitemShut {NoStop}%
\bibitem [{\citenamefont {Kresse}\ and\ \citenamefont
  {Furthm{\"u}ller}(1996)}]{Kresse1996PRB}%
  \BibitemOpen
  \bibfield  {author} {\bibinfo {author} {\bibfnamefont {G.}~\bibnamefont
  {Kresse}}\ and\ \bibinfo {author} {\bibfnamefont {J.}~\bibnamefont
  {Furthm{\"u}ller}},\ }\bibfield  {title} {\bibinfo {title} {{Efficient
  iterative schemes for ab initio total-energy calculations using a plane-wave
  basis set}},\ }\href {https://doi.org/10.1103/PhysRevB.54.11169} {\bibfield
  {journal} {\bibinfo  {journal} {Physical Review B}\ }\textbf {\bibinfo
  {volume} {54}},\ \bibinfo {pages} {11169} (\bibinfo {year}
  {1996})}\BibitemShut {NoStop}%
\bibitem [{\citenamefont {Kresse}\ and\ \citenamefont
  {Joubert}(1999)}]{Kresse1999PRB}%
  \BibitemOpen
  \bibfield  {author} {\bibinfo {author} {\bibfnamefont {G.}~\bibnamefont
  {Kresse}}\ and\ \bibinfo {author} {\bibfnamefont {D.}~\bibnamefont
  {Joubert}},\ }\bibfield  {title} {\bibinfo {title} {{From ultrasoft
  pseudopotentials to the projector augmented-wave method}},\ }\href
  {https://doi.org/10.1103/PhysRevB.59.1758} {\bibfield  {journal} {\bibinfo
  {journal} {Physical Review B}\ }\textbf {\bibinfo {volume} {59}},\ \bibinfo
  {pages} {1758} (\bibinfo {year} {1999})}\BibitemShut {NoStop}%
\bibitem [{\citenamefont {Ying}\ \emph {et~al.}(2022)\citenamefont {Ying},
  \citenamefont {Liang}, \citenamefont {Du}, \citenamefont {Zhang},
  \citenamefont {Zeng},\ and\ \citenamefont {Zhong}}]{ying2022thermal}%
  \BibitemOpen
  \bibfield  {author} {\bibinfo {author} {\bibfnamefont {P.}~\bibnamefont
  {Ying}}, \bibinfo {author} {\bibfnamefont {T.}~\bibnamefont {Liang}},
  \bibinfo {author} {\bibfnamefont {Y.}~\bibnamefont {Du}}, \bibinfo {author}
  {\bibfnamefont {J.}~\bibnamefont {Zhang}}, \bibinfo {author} {\bibfnamefont
  {X.}~\bibnamefont {Zeng}},\ and\ \bibinfo {author} {\bibfnamefont
  {Z.}~\bibnamefont {Zhong}},\ }\bibfield  {title} {\bibinfo {title} {{Thermal
  transport in planar sp2-hybridized carbon allotropes: A comparative study of
  biphenylene network, pentaheptite and graphene}},\ }\href
  {https://doi.org/10.1016/j.ijheatmasstransfer.2021.122060} {\bibfield
  {journal} {\bibinfo  {journal} {{International Journal of Heat and Mass
  Transfer}}\ }\textbf {\bibinfo {volume} {183}},\ \bibinfo {pages} {122060}
  (\bibinfo {year} {2022})}\BibitemShut {NoStop}%
\bibitem [{\citenamefont {Li}\ \emph {et~al.}(2019)\citenamefont {Li},
  \citenamefont {Xiong}, \citenamefont {Sievers}, \citenamefont {Hu},
  \citenamefont {Fan}, \citenamefont {Wei}, \citenamefont {Bao}, \citenamefont
  {Chen}, \citenamefont {Donadio},\ and\ \citenamefont
  {Ala-Nissila}}]{li2019influence}%
  \BibitemOpen
  \bibfield  {author} {\bibinfo {author} {\bibfnamefont {Z.}~\bibnamefont
  {Li}}, \bibinfo {author} {\bibfnamefont {S.}~\bibnamefont {Xiong}}, \bibinfo
  {author} {\bibfnamefont {C.}~\bibnamefont {Sievers}}, \bibinfo {author}
  {\bibfnamefont {Y.}~\bibnamefont {Hu}}, \bibinfo {author} {\bibfnamefont
  {Z.}~\bibnamefont {Fan}}, \bibinfo {author} {\bibfnamefont {N.}~\bibnamefont
  {Wei}}, \bibinfo {author} {\bibfnamefont {H.}~\bibnamefont {Bao}}, \bibinfo
  {author} {\bibfnamefont {S.}~\bibnamefont {Chen}}, \bibinfo {author}
  {\bibfnamefont {D.}~\bibnamefont {Donadio}},\ and\ \bibinfo {author}
  {\bibfnamefont {T.}~\bibnamefont {Ala-Nissila}},\ }\bibfield  {title}
  {\bibinfo {title} {{Influence of thermostatting on nonequilibrium molecular
  dynamics simulations of heat conduction in solids}},\ }\href
  {https://doi.org/10.1063/1.5132543} {\bibfield  {journal} {\bibinfo
  {journal} {The Journal of Chemical Physics}\ }\textbf {\bibinfo {volume}
  {151}},\ \bibinfo {pages} {234105} (\bibinfo {year} {2019})}\BibitemShut
  {NoStop}%
\bibitem [{\citenamefont {Thomas}\ \emph {et~al.}(2010)\citenamefont {Thomas},
  \citenamefont {Turney}, \citenamefont {Iutzi}, \citenamefont {Amon},\ and\
  \citenamefont {McGaughey}}]{Thomas_PhysRevB_2010}%
  \BibitemOpen
  \bibfield  {author} {\bibinfo {author} {\bibfnamefont {J.~A.}\ \bibnamefont
  {Thomas}}, \bibinfo {author} {\bibfnamefont {J.~E.}\ \bibnamefont {Turney}},
  \bibinfo {author} {\bibfnamefont {R.~M.}\ \bibnamefont {Iutzi}}, \bibinfo
  {author} {\bibfnamefont {C.~H.}\ \bibnamefont {Amon}},\ and\ \bibinfo
  {author} {\bibfnamefont {A.~J.~H.}\ \bibnamefont {McGaughey}},\ }\bibfield
  {title} {\bibinfo {title} {Predicting phonon dispersion relations and
  lifetimes from the spectral energy density},\ }\href
  {https://doi.org/10.1103/PhysRevB.81.081411} {\bibfield  {journal} {\bibinfo
  {journal} {Physical Review B}\ }\textbf {\bibinfo {volume} {81}},\ \bibinfo
  {pages} {081411} (\bibinfo {year} {2010})}\BibitemShut {NoStop}%
\bibitem [{\citenamefont {Feng}\ \emph {et~al.}(2015)\citenamefont {Feng},
  \citenamefont {Qiu},\ and\ \citenamefont {Ruan}}]{Feng_2015}%
  \BibitemOpen
  \bibfield  {author} {\bibinfo {author} {\bibfnamefont {T.}~\bibnamefont
  {Feng}}, \bibinfo {author} {\bibfnamefont {B.}~\bibnamefont {Qiu}},\ and\
  \bibinfo {author} {\bibfnamefont {X.}~\bibnamefont {Ruan}},\ }\bibfield
  {title} {\bibinfo {title} {Anharmonicity and necessity of phonon eigenvectors
  in the phonon normal mode analysis},\ }\href
  {https://doi.org/10.1063/1.4921108} {\bibfield  {journal} {\bibinfo
  {journal} {Journal of Applied Physics}\ }\textbf {\bibinfo {volume} {117}},\
  \bibinfo {pages} {195102} (\bibinfo {year} {2015})}\BibitemShut {NoStop}%
\bibitem [{\citenamefont {Berendsen}\ \emph {et~al.}(1984)\citenamefont
  {Berendsen}, \citenamefont {Postma}, \citenamefont {Van~Gunsteren},
  \citenamefont {DiNola},\ and\ \citenamefont {Haak}}]{berendsen1984molecular}%
  \BibitemOpen
  \bibfield  {author} {\bibinfo {author} {\bibfnamefont {H.~J.}\ \bibnamefont
  {Berendsen}}, \bibinfo {author} {\bibfnamefont {J.~v.}\ \bibnamefont
  {Postma}}, \bibinfo {author} {\bibfnamefont {W.~F.}\ \bibnamefont
  {Van~Gunsteren}}, \bibinfo {author} {\bibfnamefont {A.}~\bibnamefont
  {DiNola}},\ and\ \bibinfo {author} {\bibfnamefont {J.~R.}\ \bibnamefont
  {Haak}},\ }\bibfield  {title} {\bibinfo {title} {{Molecular dynamics with
  coupling to an external bath}},\ }\href {https://doi.org/10.1063/1.448118}
  {\bibfield  {journal} {\bibinfo  {journal} {The Journal of Chemical Physics}\
  }\textbf {\bibinfo {volume} {81}},\ \bibinfo {pages} {3684} (\bibinfo {year}
  {1984})}\BibitemShut {NoStop}%
\bibitem [{\citenamefont {Tuckerman}(2010)}]{tuckerman2010statistical}%
  \BibitemOpen
  \bibfield  {author} {\bibinfo {author} {\bibfnamefont {M.}~\bibnamefont
  {Tuckerman}},\ }\href@noop {} {\emph {\bibinfo {title} {{Statistical
  mechanics: theory and molecular simulation}}}}\ (\bibinfo  {publisher}
  {Oxford university press},\ \bibinfo {year} {2010})\BibitemShut {NoStop}%
\bibitem [{\citenamefont {Bussi}\ and\ \citenamefont
  {Parrinello}(2007)}]{bussi2007accurate}%
  \BibitemOpen
  \bibfield  {author} {\bibinfo {author} {\bibfnamefont {G.}~\bibnamefont
  {Bussi}}\ and\ \bibinfo {author} {\bibfnamefont {M.}~\bibnamefont
  {Parrinello}},\ }\bibfield  {title} {\bibinfo {title} {{Accurate sampling
  using Langevin dynamics}},\ }\href
  {https://doi.org/10.1103/PhysRevE.75.056707} {\bibfield  {journal} {\bibinfo
  {journal} {Physical Review E}\ }\textbf {\bibinfo {volume} {75}},\ \bibinfo
  {pages} {056707} (\bibinfo {year} {2007})}\BibitemShut {NoStop}%
\end{thebibliography}
\end{document}